\documentclass[fleqn,usenatbib]{mnras}

\usepackage{newtxtext,newtxmath}
\usepackage{pifont}
\usepackage{fontawesome5}
\usepackage{bbding}
\usepackage{tablefootnote}

\usepackage[T1]{fontenc}

\DeclareRobustCommand{\VAN}[3]{#2}
\let\VANthebibliography\thebibliography
\def\thebibliography{\DeclareRobustCommand{\VAN}[3]{##3}\VANthebibliography}

\usepackage{graphicx}	
\usepackage{amsmath}	
\usepackage{longtable,booktabs}

\title[Galactic distribution of PNe with different dust types]{The Galactic distribution of planetary nebulae with different types of dust}

\author[Hern\'andez-Ju\'arez et al.]{
Diego Hern\'andez-Ju\'arez,$^{1}$\thanks{E-mail: dbhernandez@astro.unam.mx}
M\'onica Rodr\'iguez,$^{2}$
and Miriam Pe\~na$^{1}$
\\
$^{1}$Universidad Nacional Aut\'onoma de M\'exico, Instituto de Astronom\'ia,  Apdo. Postal 70264, Ciudad Universitaria, Ciudad de M\'exico, M\'exico\\
$^{2}$Instituto Nacional de Astrof\'isica, \'Optica y Electr\'onica, Luis Enrique Erro 1, Tonantzintla 72840, Puebla, M\'exico\\
}

\date{Accepted XXX. Received YYY; in original form ZZZ}

\pubyear{2026}

\begin{document}
\label{firstpage}
\pagerange{\pageref{firstpage}--\pageref{lastpage}}
\maketitle

\begin{abstract}
We identify different dust features in our compilation of infrared spectra for 267 planetary nebulae (PNe) from the {\sl Spitzer}, {\sl ISO}, and {\sl IRAS} telescopes. We classify 209 objects according to their dust type: mixed dust (MD), oxygen-rich dust (ORD), carbon-rich dust (CRD), PNe with only polycyclic aromatic hydrocarbons (PAHs) in their spectra (oPAH), and featureless (F) PNe. We study statistically the distributions of surface brightness and diameter of PNe with different types of dust as well as their distributions in the Galaxy. We find that both MD and ORD PNe are closer to the Galactic centre than CRD and oPAH PNe, and that the Galactic distributions of each pair of groups are statistically compatible, suggesting that they have similar progenitors. Since oPAH PNe have, on average, larger diameters and lower surface brightness than CRD PNe, we suggest that oPAH PNe are evolved CRD PNe. On the other hand, F PNe have the lowest surface brightness and the largest diameters, suggesting they could contain evolved PNe from any initial type of dust. Among the PNe with silicates, we find that only a few ORD PNe have just amorphous silicates in their spectra, and their distributions of Galactocentric distances and Galactic heights suggest that they had low-mass progenitors. MD PNe with both amorphous and crystalline silicates have the largest surface brightness and the smallest diameters and might be the earliest stages of PNe with the most massive and metal-rich progenitors.
\end{abstract}
\begin{keywords}
Planetary nebulae: general -- dust -- Galaxy: evolution -- local interstellar matter
\end{keywords}



\section{Introduction}\label{sec:introduction}

Asymptotic Giant Branch (AGB) stars are one of the main sources of stellar dust in the Galaxy \citep[see, e.g.,][and references therein]{Schneider2024}. AGB stars are the descendants of stars with initial masses between 1 and 8 M$_\odot$ (low- and intermediate-mass stars).  During this phase, in the outermost regions of the stellar atmosphere, the suitable conditions of pressure and temperature exist for dust grains to form and grow \citep{whittet2002dust}. The critical elements that lead to dust formation are oxygen and carbon, because they are relatively more abundant than the other elements and because they can form solid compounds. Carbon and oxygen atoms combine to produce the CO molecule, a stable molecule that is difficult to dissociate, and the type of dust that an AGB forms depends on which of these two elements remains free in the stellar atmosphere. That is, the dust chemistry in the atmosphere of AGB stars is determined by the C/O abundance ratio. If C/O~$>1$ the dust chemistry will be carbon-rich, and compounds like SiC, amorphous carbon, and Polycyclic Aromatic Hydrocarbons (PAHs) are expected to form. If C/O~$<1$, the chemistry will be oxygen-rich and lead to the formation of compounds like silicates and oxides. 

The C/O abundance ratio is determined by processes that occur during the AGB phase, the most important ones being the third dredge-up (TDU) and hot bottom burning \citep[HBB,][]{Scalo1975, Iben1981}. The TDU leads to the mixing of carbon-rich material to the outer layers of the star, whereas HBB destroys carbon and produces nitrogen at the base of the star's convective envelope. Stellar evolution models differ on the precise mass limits at which these processes occur and in their efficiency \citep{Weiss2009, Karakas2016, Ventura2018, Rees2024}. In general, the models predict that the range of progenitor masses leading to carbon-rich objects is very wide at low metallicities and gets progressively narrower as the metallicity increases. At solar metallicity, the envelopes of AGB stars with initial masses of $\sim2$--4~M$_\odot$ will end up being carbon rich, whereas stars outside this range of masses, with lower and higher initial masses, will remain oxygen rich. At low metallicities, most AGB stars will have carbon-rich envelopes and, at metallicities above solar, most objects will end up being oxygen rich. 

After the AGB phase, the star evolves to the planetary nebula (PN) phase, where the dust is preserved \citep{Stasinska1999}. The analyses of the dust composition in PNe of the Magellanic Clouds \citep{Stanghellini2007} and Galactic PNe \citep{Stanghellini2012} have shown that the predictions of  stellar evolution models are in broad agreement with the observations. \citet{Stanghellini2007, Stanghellini2012} use infrared spectra from the {\sl Spitzer Space Telescope} to classify 66 PNe of the Magellanic Clouds and 150 Galactic PNe into featureless objects (with black-body emission but no dust features), objects with carbon-rich dust, objects with oxygen-rich dust, and mixed-dust objects (with emission features attributed to silicates and PAHs). They find that at the low metallicities of the Magellanic Clouds, most of the PNe are featureless or carbon rich, with just 7 per cent of oxygen-rich PNe and no mixed-dust objects, whereas the higher metallicity of the Galaxy leads to the formation of many oxygen-rich and mixed-dust PNe, with the latter mainly located near the Galactic centre.

\citet{GarciaHernandez2014} study the physical conditions and chemical abundances and dust types of their sample of 131 Galactic PNe, and compare their results with the predictions of stellar evolution models. They conclude that mixed-dust PNe are the dominant population in the Galactic bulge and have high metallicity, massive progenitors, with masses between 3 and 5 M$_\odot$. On the other hand, they find that PNe with oxygen-rich and carbon-rich dust are mostly in the Galactic disc and have low metallicity progenitors with lower masses (1--1.5 M$_\odot$ for the oxygen-rich PNe, 1.9--3 M$_\odot$ for the carbon-rich PNe), although they mention that a few oxygen-rich PNe and a significant fraction of carbon-rich objects could have more massive progenitors. \citet{GarciaHernandez2014} also argue that featureless PNe are evolved PNe that had both high and low metallicity progenitors.

\citet{DelgadoInglada2015} explore further the relation between chemical abundances and dust types in Galactic PNe by restricting the analysis to 20 objects that have good-quality optical spectra. They confirm that PNe with carbon-rich dust mostly have intermediate masses and low metallicities and that PNe with mixed dust had progenitors with the highest masses and metallicities. However, they find that most of their sample PNe with oxygen-rich dust also had high-mass, high-metallicity progenitors, although this difference with the results of \citet{GarciaHernandez2014} could be due to the fact that the sample of \citet{DelgadoInglada2015} is small.

Some of these results are based on small samples, or depend on uncertain distance estimates, or use optical spectra of poor quality. Besides, there are stellar and nebular parameters that could be related to the mass and metallicity of the PN progenitor stars, and whose relationship with the type of dust found in the nebula has not been explored. In this paper, we present a compilation of infrared spectra for 267 PNe from the {\sl Spitzer}, {\sl ISO}, and {\sl IRAS} telescopes (\S\ref{sec:infrared spectra}). With these spectra we classify 209 PNe according to their type of dust (\S\ref{sec:classification}), obtaining a sample that is 40 per cent larger than previous ones. We compare our classification to previous ones, finding differences in some cases (\S\ref{sec:comparison_other}). We then search for relations between the PN dust type and the nebular diameter, H$\alpha$ surface brightness, and Galactic distribution (\S\ref{sec:diameter} and \S\ref{sec:distribution}). Our conclusions are presented in \S\ref{sec:conclusion}. The relations between the type of dust and other parameters, such as the nebular morphology and the spectral type of the central star, will be explored in a future paper.

\section{Compilation of infrared spectra}\label{sec:infrared spectra}

Most of our sample was compiled from the {\sl Spitzer Space Telescope} database. We used the NASA/IPAC Infrared Science Archive,\footnote{\url{https://irsa.ipac.caltech.edu/applications/Spitzer/SHA/}} where we looked for those programs centred in Galactic PNe or that contained observations for objects  classified as 'true PNe' in the HASH database (Hong Kong/AAO/Strasbourg H$\alpha$ Planetary Nebulae Database, \citealt[][]{Parker2016}). The programs we use are the following: 45 (Principal Investigator, PI: Roelling, 11 PNe), 77 (PI: Rieke, 4 PNe), 93 (PI: Cruikshank, 4 PNe), 3235 (PI: Waelkens, 2 PNe), 3633 (PI: Bobrowsky, 40 PNe), 20049 (PI: Kwitter, 1 PNe), 20590 (PI: Sahai, 2 PNe), 30430 (PI: Dinerstein, 6 PNe), 30550 (PI: Houck, 8), 30652 (PI: Bernard-Salas, 2 PNe), 40115 (PI: Fazio, 3 PNe). 40035 (PI: Bernard-Salas, 3 PNe), 50168 (PI: De Marco, 5 PNe), and 50261 (PI: Stanghellini, 131 PNe).

The spectrometer in the {\sl Spitzer Space Telescope},  the  Infrared Spectrograph (IRS), has four modules covering the wavelength range 5.2--38.0 $\mu$m \citep{Werner2004, Houck2004}. Two modules have a variable low-resolution, between 6 and 128, and cover the full wavelength range, whereas the other two modules have a resolution of 600 in the range from 9.9 to 37.2~$\mu$m.

We used the spectra taken with the four spectrograph modules of {\sl Spitzer}, removing bad pixels and rescaling parts of the spectra when necessary. We also added the two spectra available for each object in the {\sl Spitzer} database. These corrections were made with the \texttt{SMART} program \citep{Hidgon2004}. The spectra of the low resolution modules have available a sky-subtracted version. In order to avoid background contamination, we use these sky-subtracted spectra to look for the PAH features, although background contamination is not expected to be important in most cases \citep{Stanghellini2012}. When the higher resolution spectra are available, we use them to look for the other dust features. We show in Appendix~\ref{app:spectra} the {\sl Spitzer} spectra that, as far as we know, have not been previously published.

To obtain the largest possible sample, we also searched for published spectra from the {\sl ISO} Short Wave Spectrometer (SWS) \citep{Kessler2002} and {\sl IRAS} Low Resolution Spectrometer (LRS) \citep{Helou1995}. {\sl ISO}/SWS covered the 2.4--45~$\mu$m band with spectral resolutions in the range 1000--35000. {\sl IRAS}/LRS covered the 7.7 to 22.7~$\mu$m wavelength range with a spectral resolution of 20--60. We only looked for {\sl  ISO} and {\sl IRAS} spectra for  objects that have no {\sl Spitzer} spectra, and we only used works that show the observed spectra in figures  where we can identify or confirm the presence or absence of the dust features.

In total, we have compiled spectra for 267 PNe: 228 from {\sl Spitzer}, 10 from {\sl ISO}, 15 from {\sl IRAS}, 5 with spectra from {\sl IRAS} and {\sl ISO}, 8 that have different parts of the spectrum observed by {\sl Spitzer} and {\sl IRAS}, and 1 that has different parts of the spectrum observed by {\sl Spitzer} and {\sl ISO}. The spectra from {\sl IRAS} have the most limited wavelength coverage, 7.7--22.7 $\mu$m \citep{Emerson1985}. Besides, for those objects with {\sl Spitzer} spectra, 7 do not have data between 5.2 and 19.9~$\mu$m, and 9 objects only have data for this wavelength range.

More than 50 per cent of our sample comes from the {\sl Spitzer} program 50261, which searched for compact PNe (with angular diameters below 4\arcsec). For this reason, a large part of our sample is likely to be composed of young PNe, although some compact PNe can be old and distant \citep{Stanghellini2012}. Besides, a significant part of our sample comes from works centred in the study of PNe with mixed-chemistry dust \citep[as the objects from][]{PereaCalderon2009}. These sample characteristics may introduce biases that should be considered when analysing the results.

\section{Dust features and PNe classification}\label{sec:classification}

In all the compiled infrared spectra from {\sl Spitzer},  {\sl ISO}, and {\sl IRAS}, we search for features of PAHs, silicon carbide (SiC), amorphous and crystalline silicates, and a broad feature around 30 \micron. These are the most easily identified features in the infrared spectrum region we are analysing and have been frequently studied.

\begin{figure}
	\includegraphics[width=\columnwidth]{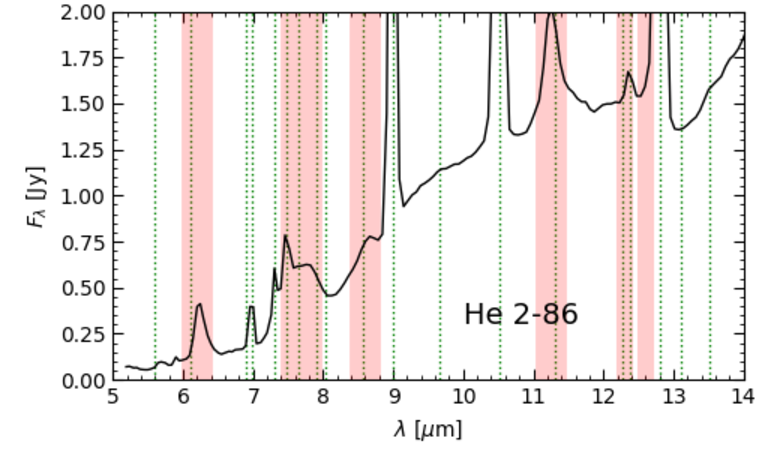}
    \caption{Example of a spectrum with PAH features, whose expected positions are indicated with pink stripes. The green dotted lines show the wavelengths of nebular emission lines: \ion{Mg}{V}] 5.60 \micron, H$_2$ S(6) 6.10 \micron, H$_2$ S(5) 6.90 \micron, \ion{He}{I} 6.99 \micron, [\ion{Na}{III}] 7.31 \micron, \ion{H}{I} 7.48 \micron, [\ion{Ne}{IV}] 7.65 \micron, [\ion{Ar}{V}] 7.90 \micron, H$_2$ S(4) 8.03 \micron, [\ion{Ne}{IV}] 8.56 \micron, H$_2$ S(3)  9.66 \micron, [\ion{S}{IV}] at 10.51 \micron, \ion{H}{I} 11.30 \micron, H$_2$ S(2) 12.28 \micron, \ion{He}{I} 12.39 \micron, [\ion{Ne}{II}] 12.82 \micron, [\ion{Ar}{V}] 13.10 \micron, and [\ion{Mg}{V}] 13.52 \micron.}
    \label{fig:PAH}
\end{figure}

In the wavelength range that we are considering, features that are usually attributed to PAHs can be detected at 6.2, 7.7, 8.6, 11.2, 12.3, and 12.6 \micron\ \citep{Allamandola1989}. Most of these features can be clearly seen in the spectrum of the PN Hen\,2-86, which we show in Fig.~\ref{fig:PAH} for illustrative purposes. The approximate positions of these PAH features are marked with pink stripes in the figure. 

\begin{figure}
	\includegraphics[width=\columnwidth]{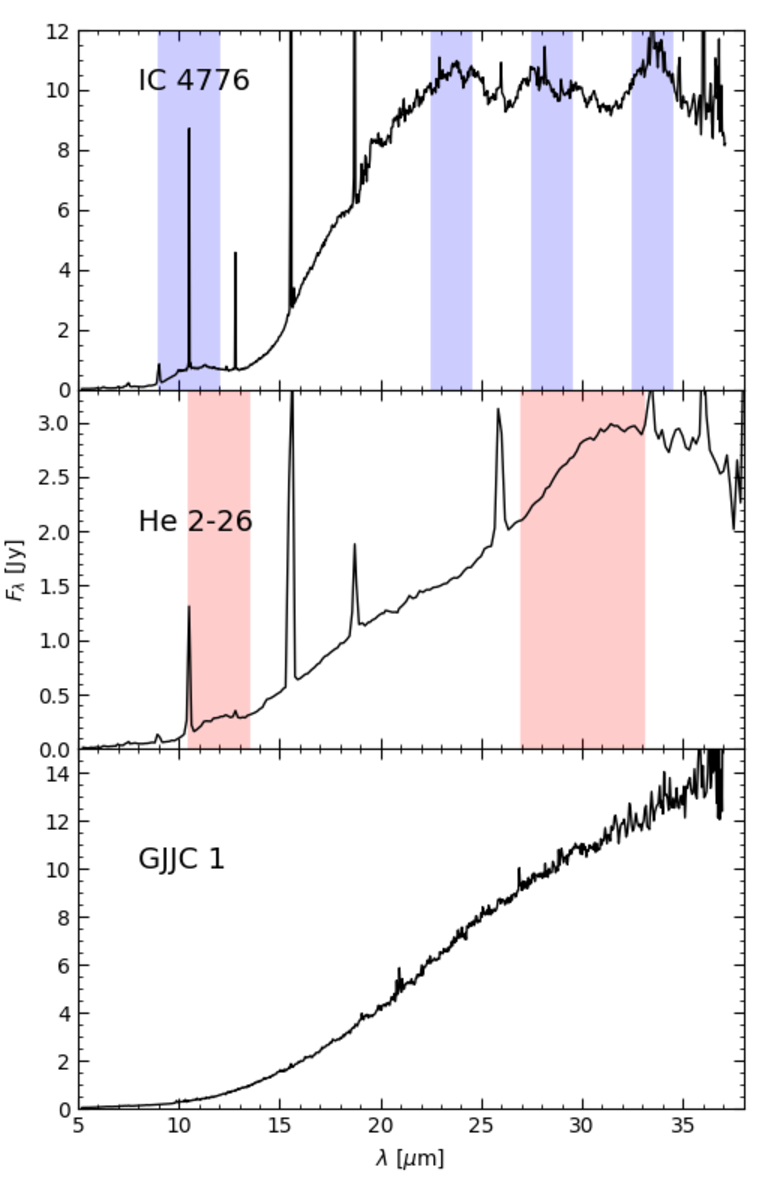}
    \caption{Top panel: spectrum of IC\,4776, which shows features of amorphous silicates at $\sim10$ \micron\ and  crystalline silicates at $\sim23$, 28 and 33 \micron, as indicated with the blue stripes. Middle panel: spectrum of Hen\,2-26 showing the SiC feature at around 11 \micron\ and the 30 \micron\ feature, both marked with pink stripes. Bottom panel: spectrum of GJJC\,1, a PN with no apparent dust features. }
    \label{fig:T}
\end{figure}

The most prominent feature due to amorphous silicates is detected at $\sim10$~\micron. There is another feature due to amorphous silicates at $\sim18$~\micron, but we do not use it for classification purposes because two parts of the {\sl Spitzer} spectrum are spliced together in this region. The features of crystalline silicates are located at around 23, 28, and 33 \micron. All these silicate features are illustrated for the PN IC\,4776 in Fig.~\ref{fig:T} (top panel).

The SiC feature can be seen at around 11 \micron, and there is another feature associated with carbon-rich environments at 30 \micron. Some suggestions of compounds that could explain the 30 \micron\ feature are magnesium sulfide \citep[MgS,][]{Goebel1985, Nuth1985}, hydrogenated amorphous carbon \citep{Grishko2001}, and molecules similar to PAHs \citep{Papoular2011}. The features of SiC and 30 \micron\ are shown for the PN Hen\,2-26 in the middle panel of Fig.~\ref{fig:T}.

Some PNe do not have any clear dust features in their spectrum, although they do show blackbody-like emission, most probably due to dust emission. The bottom panel of Fig.~\ref{fig:T} shows an example of this kind of featureless spectra.

The sample PNe are classified into five groups depending on the dust features present in their spectra: (i) oxygen-rich dust (ORD) PNe, with amorphous silicates and/or crystalline silicates; (ii) mixed-dust (MD) PNe, with PAHs plus amorphous silicates and/or crystalline silicates; (iii) carbon-rich dust (CRD) PNe, with SiC and/or the 30 \micron\ feature, with or without PAHs; (iv) only PAH (oPAH) PNe, with PAHs and no other dust features;  and (v) featureless (F) PNe, with no dust features.
These groups are similar to those defined by other works, with the exception of the oPAH group, which was included in the CRD group in previous analyses \citep[][]{Stanghellini2012, GarciaHernandez2014}. We define here the oPAH PNe as a separate group because PAHs are found both in CRD and MD PNe \citep{Gorny2001, GuzmanRamirez2012}.

From our original sample of 267 PNe with available infrared spectra, 221 have a full spectrum in the range from 5 to 38 \micron\ and most of them, 209 PNe, can be classified into the dust groups defined above, whereas the remaining 12 PNe have low signal-to-noise spectra that prevents their classification, although their spectra can be used to verify the presence or absence of some dust features. On the other hand, two objects that have incomplete spectra Hb\,5 and Hu\,2-1, can be classified as CRD PNe because they have the SiC feature or the 30 \micron\ feature. Therefore, the sample that can be classified following the procedure described above contains 209 PNe with a well-defined type of dust. Table~\ref{tab:Dust} shows the numbers and percentages of PNe in each group and subgroup. This sample is $\sim40$ per cent larger than the largest sample previously analysed \citep[][with 150 PNe]{Stanghellini2012}. 

\begin{table}
	\centering
	\caption{Numbers of PNe with different types of dust. The percentages in the second column are for the full sample, whereas the percentages in the last column are for each type of dust.}
	\label{tab:Dust}
	\begin{tabular}{lllr} 
		\hline
		Type & N (\%) & Features & N (\%) \\
		\hline
         MD	& 73 (35 \%) & Crystalline silicates, PAHs &  51 (70 \%) \\
         	&    		 & Am. and crys. silicates, PAHs & 22 (30 \%) \\
         	&	   		  & 				    &    \\
        ORD	& 53 (25 \%) & Amorphous silicates  & 13 (25 \%) \\
         	&	    	 & Crystalline silicates & 24 (45 \%) \\
        	        &       	 & Am. and crys. silicates & 16 (30 \%) \\
         	&	 		 & 				    &    \\
        CRD$^a$ & 36 (17 \%) & SiC, can have PAHs  & 10 (28 \%) \\
         	&	   	 	 &  30 \micron, can have PAHs & 8 (22 \%) \\
                 &       	 & SiC and 30 \micron, can have PAHs & 18 (50 \%) \\
         	&	       &				    &    \\
       oPAH & 12 (6 \%)  &  PAHs &  \\
        &  &  &  \\
        F  &  35 (17 \%)  & No features &  \\
         &   & &  \\
        Total &  209 (100 \%)  & &  \\
		\hline
	\end{tabular}\\
	 {\raggedright   $^a$ There are 17 CRD PNe with PAH features (49 \%): 3 of them show the SiC feature, 2 show the 30 \micron\ feature, and 12 have both SiC and the 30 \micron\ feature in their spectra. \par}
\end{table}

Of the initial sample of 267 PNe, 58 objects could not be classified because they had incomplete or noisy spectral information, but we have compiled some information on the presence or absence of dust features in their spectra. Thus, information for a total of 267 PNe has been gathered. The number of PNe that have the different features is given in Table \ref{tab:Features}, where the first column indicates the feature; the second column gives the number of PNe where the spectral range in which we can see the features is available; and the third column lists the number of PNe that show the feature and the percentage of PNe with the feature. The most common features are those attributed to PAHs and silicates, while the SiC and 30 \micron\ features are less frequent. Among the silicates, most of them are crystalline.

The features identified for each object and the types of dust assigned to them are presented in Table~\ref{tab:dust_type} in Appendix~\ref{app:dust}, along with the references for their published spectra. 

	\begin{table}
    	\centering
    	\caption{PNe with specific dust features. In the percentage indicated in the last column, we only consider those PNe with data in the spectral range where we can see the feature.}
    	\begin{tabular}{llr}
      		\hline
       		Feature & PNe with data & PNe with the feature   \\
        	\hline
      		SiC  & 227 & 28 (12 \%)  \\
       		30 \micron & 233 & 26 (11 \%) \\
       		PAHs  & 227 & 100 (44 \%) \\
       		Amorphous silicates & 227 & 50  (22 \%) \\
      		Crystalline silicates & 233 & 115 (49 \%) \\
       		\hline
    	\end{tabular}  
        \label{tab:Features}
	\end{table}

\section{Comparison with previous classifications}\label{sec:comparison_other}

The largest samples of PNe classified into different dust types are those presented by \citet{Stanghellini2012} and \citet{GarciaHernandez2014}. \citet{Stanghellini2012} present {\sl Spitzer} spectra for 157 PNe and classify 150 of them into four groups, three of them similar to those of this work (MD, ORD, and F) and one that includes the objects that we classify as CRD and oPAH PNe here. \citet{GarciaHernandez2014} include 148 PNe in their classification, and follow a classification scheme similar to that of \citet{Stanghellini2012}.

Our classification contains 209 objects, including the PNe studied by \citet{Stanghellini2012} and \citet{GarciaHernandez2014}, and there are differences in the classification of some objects. We have 14 PNe with a different dust type than the one assigned by \citet{Stanghellini2012} and 16 PNe with a different type than the one assigned by \citet{GarciaHernandez2014}. An example of this is Al\,2-E (PN\,G359.3+03.6), in whose spectrum we do not observe crystalline silicate features, but \cite{Stanghellini2012} and \citet{GarciaHernandez2014} do. We list the objects with different classification in Table~\ref{tab:different_clasification_S12_GH14} in Appendix~\ref{app:differences}. The number of objects with a different classification is small, and our results in the following sections are not affected by these objects. 

In the case of the {\sl ISO} and {\sl IRAS} spectra, all of which have already been published by other authors, some of these authors also identify the dust features, and these identifications agree with ours.

\section{Nebular diameter and surface brightness}\label{sec:diameter}

\citet{GarciaHernandez2014} argue that F PNe are more evolved than other types of PNe since they have lower electron densities and higher degrees of ionization. Here we explore the evolutionary status of PNe with different types of dust using their surface brightness and nebular diameters, since these parameters can also be expected to change during the lifetime of a PN.

We use the values of intrinsic H$\alpha$ surface brightness calculated by \cite{Frew2016}, which are available for 63 per cent of our sample with a well-defined dust type. As for the diameters, we calculate their values using the angular diameters listed in the HASH database \citep{Parker2016} and the most reliable distances from \citet{HernandezJuarez2024}. These distances are based on more than one distance estimate \citep[cases A, B, and C in][]{HernandezJuarez2024}. We can calculate the diameter for 70 per cent of our sample with a well-defined dust type.

\begin{figure*}
	\includegraphics[width=1.8\columnwidth]{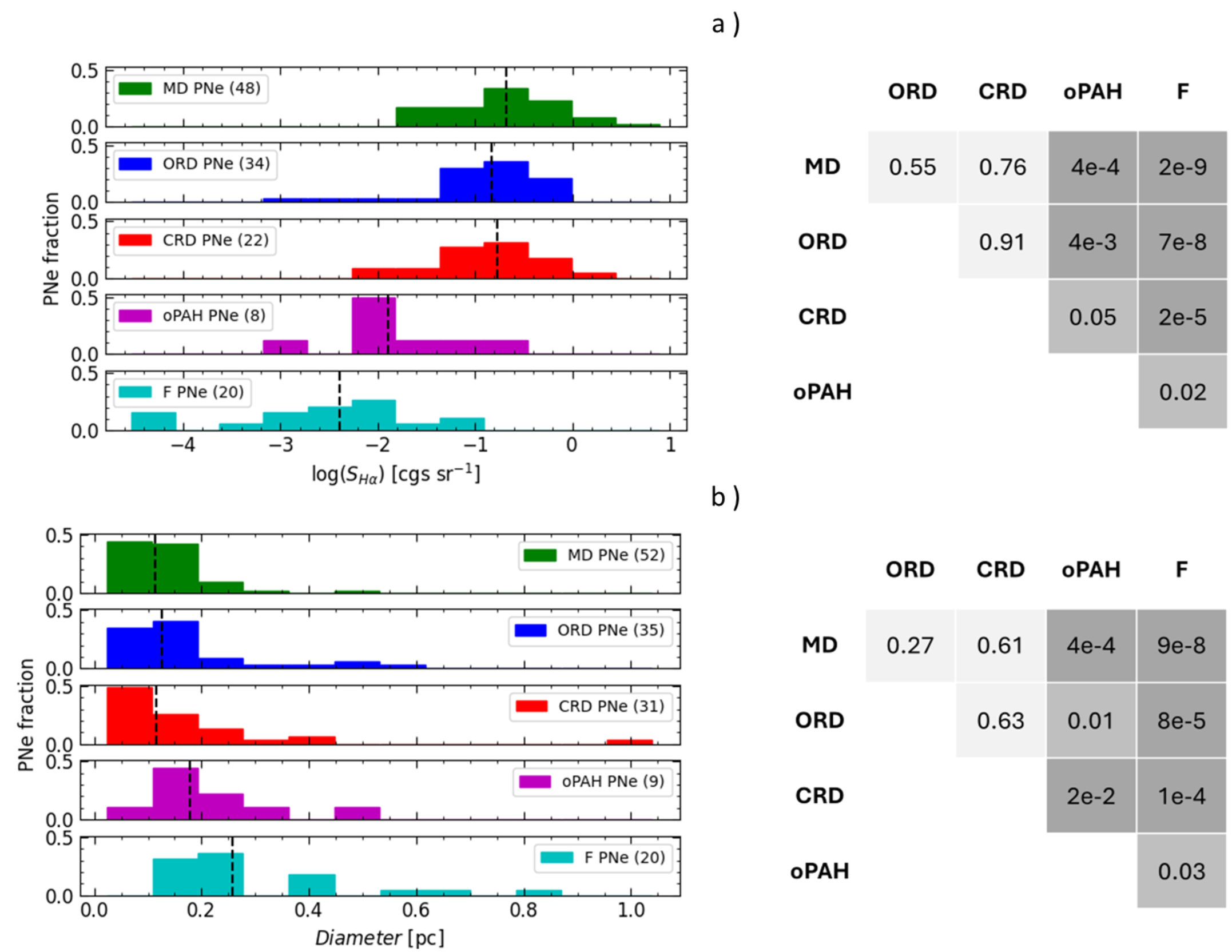}
    \caption{Distributions (left panels) and $p$-values (right panels) of:  a) the H$\alpha$ surface brightness of PNe, and b) the PN diameters, for different types of dust. The dashed lines in the left panels indicate the medians of the distributions. The right panels show the results of Kolmogorov-Smirnov (K-S) tests between each pair of dust types, with the lightest cells indicating similar distributions, and the darkest cells indicating significantly different distributions.}
    \label{fig:SHA_diam}
\end{figure*}

The left panels of Fig.~\ref{fig:SHA_diam} show the distributions of both parameters, H$\alpha$ surface brightness and nebular diameter, for PNe with different types of dust, with the numbers in parentheses indicating the sample sizes. The medians of the distributions are indicated with the dashed lines. We can see in Fig.~\ref{fig:SHA_diam} that the oPAH and, even more so, the F PNe, have lower brightness and larger diameters in average than the MD, CRD, and ORD PNe. 

The statistical significance of the similarities and differences between the distributions shown in Fig.~\ref{fig:SHA_diam} can be studied with Kolmogorov-Smirnov tests \citep[K-S,][]{press2007numerical}. The results from these tests are the $p$-values, the probabilities that the differences in the cumulative distributions of two samples can be explained by statistical fluctuations if the two samples had the same original distribution. The $p$-values of the one-dimensional K-S tests  that compare each possible pair of distributions are listed in the right panels of Fig.~\ref{fig:SHA_diam}. A large $p$-value indicates that the two distributions are compatible (they might have the same original distribution or the differences are too small to be detected with the available data); a small $p$-value indicates that there are significant differences between the samples. 

The one-dimensional K-S tests that we are using require that the effective number of data for the two distributions to be compared, $N_{\rm e}=N_1N_2/(N_1+N_2)$, is larger or equal to 4, with $N_1$ and $N_2$ the sizes of the two distributions. This condition holds for all the one-dimensional K-S tests presented in this work. The lowest values of $N_{\rm e}$ are found for the comparisons beween oPAH and CRD PNe, with $N_{\rm e}=5$--9.

As we can see from the $p$-values in Fig.~\ref{fig:SHA_diam}, MD, ORD, and CRD PNe have similar distributions of surface brightness and diameters, oPAH PNe have significantly lower surface brightness and larger diameters that these types, and F PNe have on average the lowest surface brightness and largest diameters of all dust types. This suggests that PNe lose their infrared dust features as they evolve and that oPAH could be evolved nebulae that were originally MD or CRD PNe whereas F PNe  are even more evolved nebulae that were originally MD, ORD, CRD, or oPAH PNe. The connections between these types of dust are further explored in the next section, although we should also keep in mind that oPAH and F PNe might contain objects that had different progenitors from those of the other dust types (less massive, or with lower metallicity, or with C/O~$\sim1$).

\section{Galactic distribution}\label{sec:distribution}

\begin{figure*}
	\includegraphics[width=2\columnwidth]{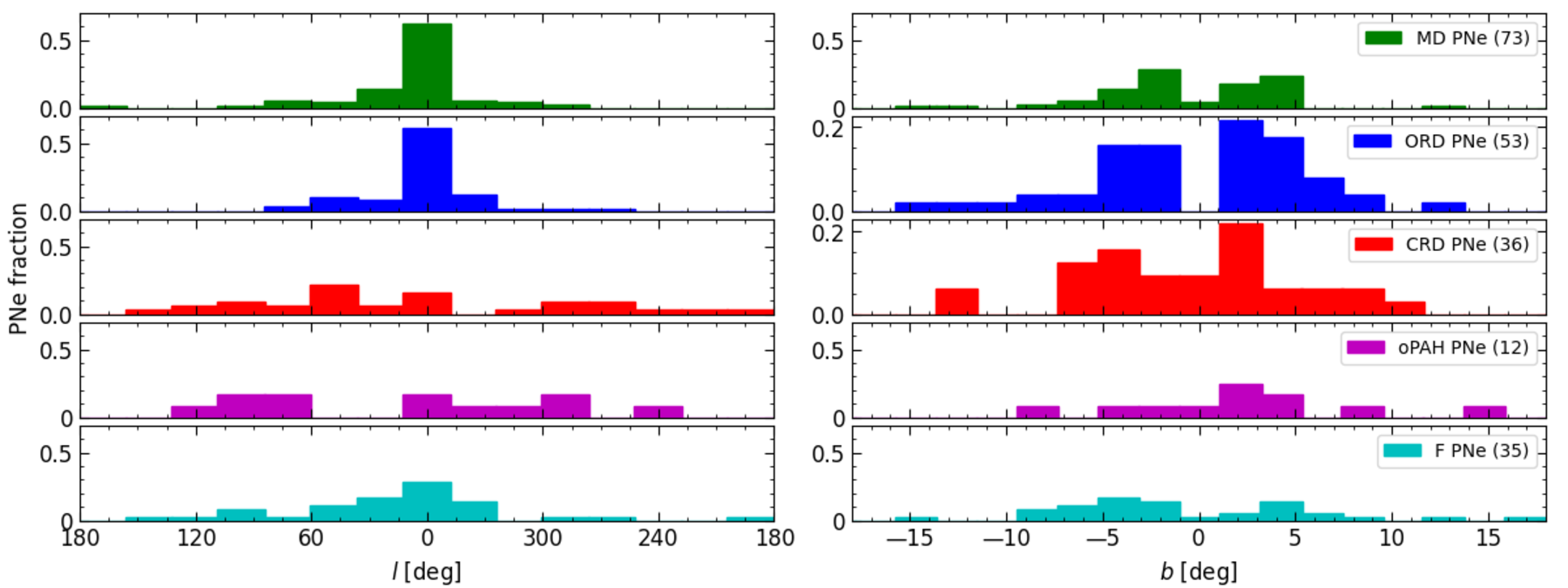}
    \caption{Distributions of Galactic longitude ($l$) and latitude ($b$) of PNe with different types of dust. The numbers in the parentheses indicate the sample sizes. Five PNe lie outside the range plotted for $b$: MaC\,2-1 (PN\,G205.8$-$26.7, CRD), NGC\,6210 (PN\,G043.1+37.7, ORD), BoBn\,1 (PN\,G108.4$-$76.1, oPAH), NGC\,246 (PN\,G118.8$-$74.7, F), and NGC\,7094 (PN\,G066.7$-$28.2, F).}
    \label{fig:longlat}
\end{figure*}

Fig.~\ref{fig:longlat} shows the distributions in the sky, in Galactic coordinates, of PNe with different types of dust. Most MD and ORD PNe can be seen to be located preferentially towards the Galactic center, whereas CRD and oPAH PNe have wider distributions in Galactic longitude. This is consistent with the Galactic distribution of oxygen-rich and carbon-rich AGB stars \citep[see][and references therein]{Lewis20} if we assume that the progenitors of MD and ORD PNe were oxygen-rich AGB stars whereas the progenitors of CRD and oPAH PNe were carbon-rich AGB stars.

On the other hand, the distributions of Galactic latitudes in Fig.~\ref{fig:longlat} drop at latitudes close to zero, especially for MD, ORD, and F PNe. Since these are also the PNe that are located mostly in the direction to the Galactic centre, this result indicates that we are missing many MD, ORD, and F PNe which are located close to the Galactic plane.

\subsection{Heliocentric distances}

In order to explore the Galactic distribution of our sample PNe, we need their distances. We have good distance estimates for 185 PNe of the sample of 209 PNe with a well-defined type of dust. This is nearly double the sample of 92 Galactic PNe used by \citet{Stanghellini2012} to study the Galactic distribution of PNe with different types of dust. As in \S\ref{sec:diameter}, the distances we use are the most reliable ones of those calculated by \citet{HernandezJuarez2024}. One of the sample PNe, BoBn\,1 (PN\,G108.4$-$76.1, an oPAH PN), belongs to the Galactic halo \citep{Quireza2007}, and we have excluded it from the analyses that follow. Therefore, we can study the Galactic distribution of 89 per cent of our sample with a well-defined type of dust.

\begin{figure*}
	\includegraphics[width=1.8\columnwidth]{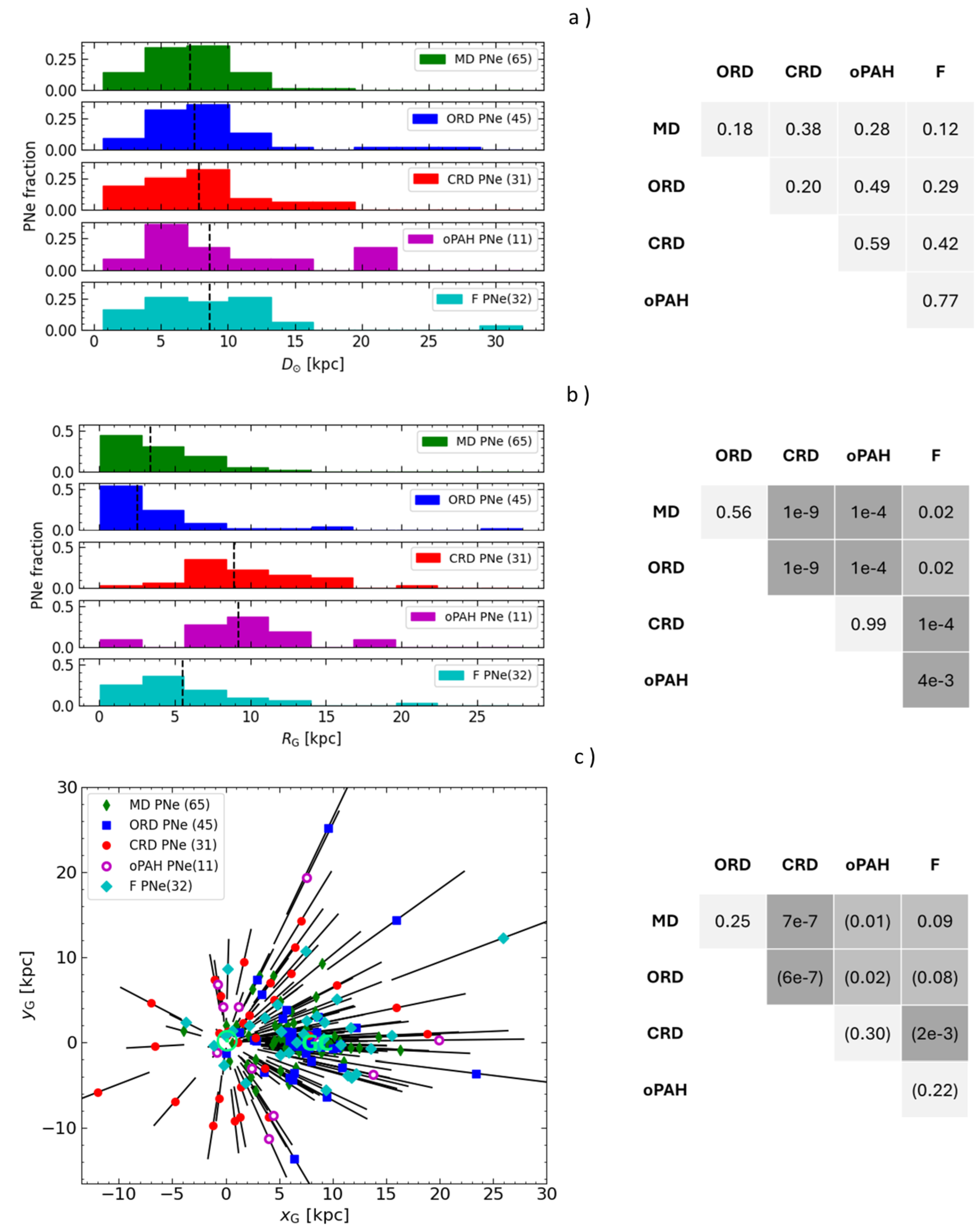}
    \caption{Distributions (right panels) and $p$-values (left panels) of a) heliocentric distances; b) Galactocentric distances; and c) projection on the Galactic plane of PNe with different types of dust. The dashed lines in the top and middle left panels indicate the medians of the distributions. In the bottom left panel, the positions of the Sun (0, 0) and the Galactic centre (8, 0) are marked in green as "$\odot$" and "GC", and the lines show the distance uncertainties. The right panels show the $p$-values obtained from one dimensional K-S tests that compare each pair of distributions. The darkest cells in these panels indicate significant differences between the distributions. The $p$-values that are based on an insufficient number of data are enclosed in parentheses}.
    \label{fig:TT}
\end{figure*}

Since the distance uncertainties are larger for the most distant sources, if the different types of dust have different distributions of heliocentric distances, this can introduce biases in our results. Hence, we first examine the heliocentric distance distribution of each type of PNe. The left panel of Fig.~\ref{fig:TT}a shows these distributions, whereas the right panel shows the $p$-values obtained by comparing each pair of distributions with one-dimensional K-S tests. The $p$-values are all larger than 0.14, implying that there are no significant differences in the distributions of heliocentric distances of PNe with different types of dust.

\subsection{Galactic distribution of PNe with different types of dust}

The middle and bottom left panels of Fig.~\ref{fig:TT} show the distributions of Galactocentric distances, $R_\mathrm{G}$, and the projections on the Galactic plane of PNe with different types of dust. The plane of the Galaxy is shown as seen from positive Galactic height and centred on the Sun. The position of the Galactic center (GC in the figure) is also indicated, and the distance uncertainties are shown with lines.

The middle and bottom right panels of Fig.~\ref{fig:TT} show the $p$-values that result when we compare each pair of distributions using K-S tests \citep{Fasano1987, press2007numerical}. The K-S tests confirm that the similarities and differences that can be seen in the middle and bottom left panels of Fig.~\ref{fig:TT} are statistically significant. As previously mentioned, the one-dimensional K-S tests are valid when the effective number of data for the two distributions to be compared, $N_{\rm e}$, is larger or equal to 4. On the other hand, the two-dimensional K-S tests used to compare the distributions of PNe in the Galactic plane require $N_{\rm e}\geq20$ and this condition is valid only for the comparisons between the MD PNe and the ORD, CRD, and F PNe (the $p$-values that are based on an insufficient number of data are enclosed in parentheses in the bottom right panel of Fig.~\ref{fig:TT}). However, the $p$-values derived from the two-dimensional K-S tests that compare the distributions on the Galactic plane are mostly in agreement with those derived from the one-dimensional K-S tests that compare the distributions of Galactocentric distances. We only show them for comparison purposes, and the conclusions presented below are based on the $p$-values implied by the one-dimensional K-S tests, all of which are statistically valid.

The results presented in Fig.~\ref{fig:TT} indicate that MD and ORD PNe are more abundant towards the Galactic center than CRD, oPAH, and F PNe. Previous works also find that many MD PNe are close to the Galactic centre, but conclude that ORD PNe have distributions intermediate between those of MD and CRD objects \citep{Stanghellini2012, GarciaHernandez2014}. The differences among the results might be due to the procedure followed by \citet{GarciaHernandez2014} to assign PNe to different parts of the Galaxy or to the use of \citet{Stanghellini2012} of a smaller sample with no statistical comparisons between the distributions. Indeed, the $p$-values shown in Fig.~\ref{fig:TT} indicate that MD and ORD PNe are distributed in a very similar way in the Galaxy.

It can also be seen in Fig.~\ref{fig:TT} that oPAH and CRD PNe also have very similar distributions. So, each pair of groups seems to have a common origin. Besides, the distributions of MD and ORD PNe are significantly different from those of oPAH and CRD PNe, whereas F PNe have intermediate distributions that in most cases are significantly different from the other dust types. Taking into account the results found in \S\ref{sec:diameter}, we conclude that most oPAH PNe are likely to be evolved CRD PNe, whereas F PNe can be even more evolved PNe that originally belonged to any of the other dust types. The groups oPAH and F can also contain or be composed of objects where a low metallicity, or a C/O abundance ratio close to one, prevented the formation of the dust compounds that we are studying. However, at least in the case of oPAH PNe, their initial masses and metallicities should not be very different from those of CRD PNe, because the distributions in the Galaxy of the two groups are very similar.

Evolved PNe might lose their emission in a dust feature if the dust grains become too cold to emit, or if they are altered or destroyed. For example, some compounds, such as fullerenes or graphite, might form on the surface of SiC grains and prevent us from observing the SiC feature \citep{Bernal2019}.
 
The similarities between the distributions of MD and ORD PNe suggests that MD PNe were originally ORD objects and some process led to the formation of PAHs in their vicinity. There are several scenarios that can explain the formation of PAHs in an environment that contains silicates \citep[see, e.g.,][and references therein]{PereaCalderon2009, GuzmanRamirez2011}, such as a very late thermal pulse that turns an oxygen-rich outflow into a carbon-rich one \citep{Waters1998}, or the dissociation of CO in a dense torus irradiated by the central star \citep{Matsuura2004, Cernicharo2004, GuzmanRamirez2011, Cox2016}.

The increase of MD and ORD PNe towards the Galactic centre suggest that these objects arise more easily from progenitors with higher metallicity, as predicted by most models of stellar evolution \citep[see, e.g.,][]{2014MNRAS.445..347K}. Previous results based on nitrogen and helium abundances suggest that the highest progenitor masses should be found among MD PNe, followed by ORD PNe \citep{GarciaHernandez2014, DelgadoInglada2015}. Besides, although some ORD PNe have been found in the Magellanic Clouds, no MD PNe have been found in these low-metallicity galaxies. Hence, we suggest that only the most massive and metal-rich oxygen-rich progenitors can form MD PNe. \cite{PereaCalderon2009} argue that dual-dust chemistry is related to low-mass progenitors, but this conclusion seems to be based on the assumption that that all bulge PNe, where this phenomenon is prevalent, had old low-mass progenitors. Since there is a significant population of young stars in the Galactic bulge \citep[see][and references therein]{2020ApJ...901..109H}, this might not be the case.

We do not explore the distributions of distances to the Galactic plane for the different types of dust because our sample is not big enough in the solar neighbourhood (see Fig.~\ref{fig:TT}), because the different types of dust have different Galactic distributions, and because we are missing many oxygen-rich PNe located close to the Galactic plane in the direction of the Galactic centre
(see Fig.~\ref{fig:longlat}).

\subsection{Galactic distribution of different dust features}

We have explored the Galactic distributions and the distributions in surface brightness and diameters of the different dust features that we identify in the spectra. In most cases, these distributions reflect the distributions of the PN dust types that contain the feature, and do not provide new insights into the origins of the PN hosting the feature. The exceptions are the distributions of amorphous and crystalline silicates in MD and ORD PNe, and below we discuss only the results concerning these features that have any statistical significance (with $p$-values of a few per cent or lower).

We divide the sample MD and ORD objects into PNe with only crystalline silicates, PNe with both crystalline and amorphous silicates, and PNe with only amorphous silicates.

\citet{Stanghellini2012} mention that ORD PNe with crystalline silicates have typically lower Galactic latitudes than ORD PNe with amorphous silicates. From this, they suggest that the progenitor stars of crystalline ORD PNe had higher masses than the progenitors of amorphous ORD PNe. Here, we revise this results using the Galactic distributions of our sample objects. We find that, of the 209 PNe classified into dust types, 13 objects show only amorphous silicates in their spectra, and all of them are ORD PNe. On average, these objects have higher Galactocentric distances and reach higher distances from the Galactic plane than the other ORD (or MD) PNe. The differences are significant with a confidence level of 92--96 per cent (i.e. $p$-values of 0.04--0.08). In agreement with \citet{Stanghellini2012}, we conclude that PNe with only amorphous silicates are ORD objects that probably had progenitors with lower masses and metallicities than PNe that also show crystalline silicates in their spectra.

\begin{figure*}
		\center
		\includegraphics[width=1.8\columnwidth]{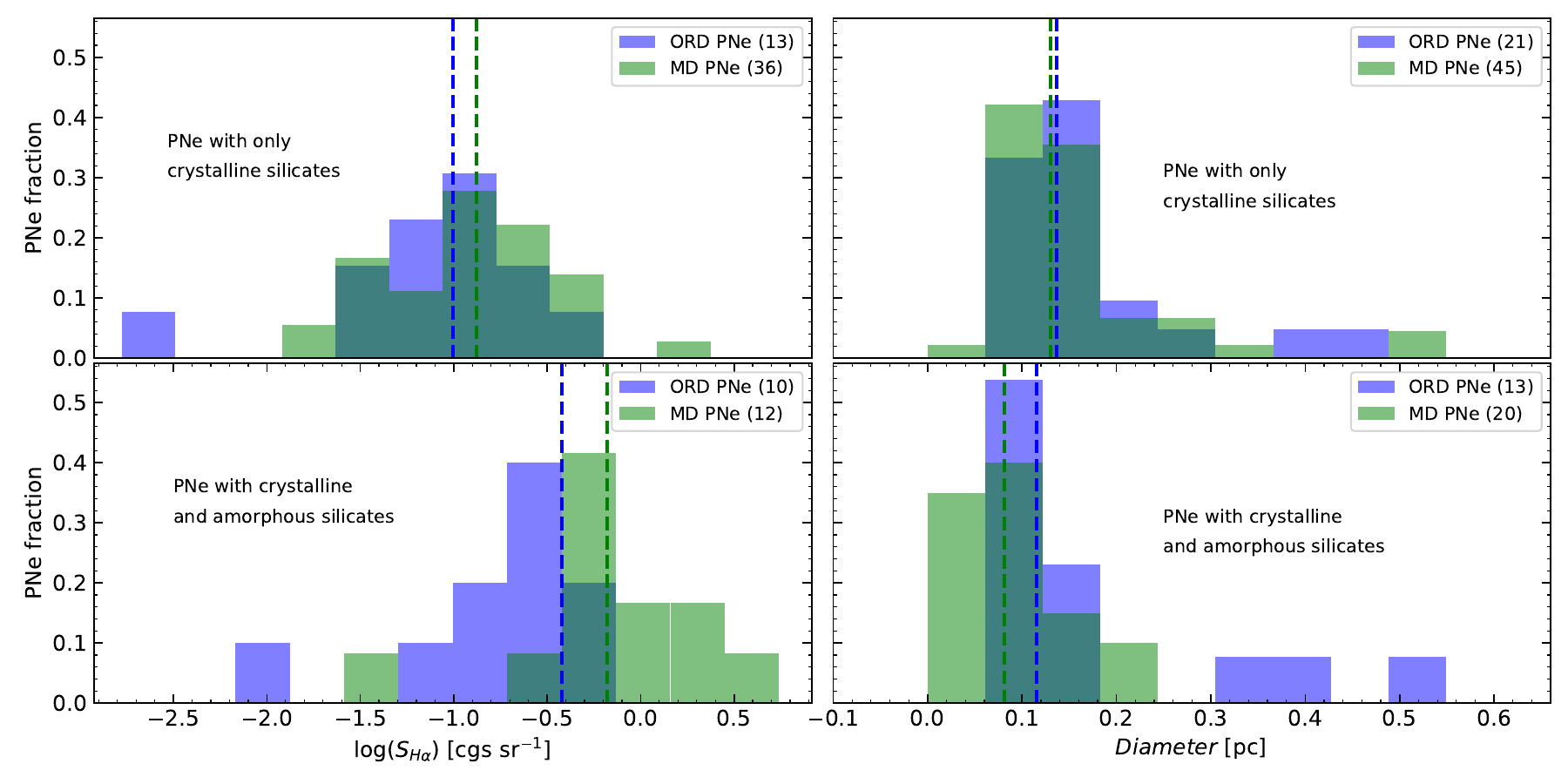}
    	\caption{Distributions of H$\alpha$ surface brightness and physical diameter for MD and ORD PNe with only crystalline silicates (top panels) and with both crystalline and amorphous silicates (bottom panels). The dashed lines indicate the medians of the distributions.}
   	 \label{fig:Sd_silicates}
	\end{figure*}

The higher metallicity near the Galactic centre can lead to higher mass-loss rates and a more efficient crystallization of silicates \citep{Waters96, Molster1999, GarciaHernandez2014}. The crystallization of silicates can occur before the PN phase \citep{Cami1998, Sylvester1999, Kemper2001, deVries2014}, but if the process continues in the PN stage, we might find that PNe with crystalline silicates are on average more evolved than PNe with amorphous silicates. The possible relationship between crystalline silicates and PNe evolution has been explored by \citet{GarciaHernandez2014}, with mixed results. We explore here the distributions of H$\alpha$ surface brightness and physical diameter for MD and ORD PNe that show different types of silicate features in their spectra. The distributions for the 9--10 PNe with only amorphous silicates that have information on the surface brightness and diameter do not show any significant difference from those of PNe with only crystalline silicates or with both types of silicate, and are not discussed further. The distributions of H$\alpha$ surface brightness and physical diameter for MD and ORD PNe with only crystalline silicates and with both crystalline and amorphous silicates are shown in Fig.~\ref{fig:Sd_silicates}.

We can see by comparing the medians of the distributions in Fig.~\ref{fig:Sd_silicates} that ORD PNe have slightly lower surface brightness and slightly larger diameters than MD PNe. As shown in \S\ref{sec:diameter}, the differences, although consistent with each other, are not statistically significant when the whole sample of MD and ORD PNe are considered. However, there are significant differences between MD and ORD PNe for those objects that show both types of silicates in their spectra, with $p$-values of 0.003 for the comparison of surface brightness and 0.02 for the comparison of diameters. Besides, MD PNe with only crystalline silicates have significantly lower surface brightness and larger diameters than MD PNe with both types of silicates, with $p$-values of 0.0001 for the two comparisons. In fact, MD PNe with both crystalline and amorphous silicates are clearly distributed towards larger surface brightness and smaller diameters than all other MD and ORD PNe. On the other hand, ORD PNe with only crystalline silicates have somewhat lower surface brightness and larger diameters than ORD PNe with both crystalline and amorphous silicates, although in this case the differences are smaller and are not statistically significant, with $p$-values of 0.1 and 0.3 respectively.

In addition to these differences in surface brightness and diameter, we know that no MD and only a few ORD PNe have been found in the low-metallicity environments of the Magellanic Clouds\footnote{\citet{Jones2017} found more ORD PNe than CRD PNe in the Large Magellanic Cloud. However, since their classification criteria is based on the presence or absence of PAH features, their ORD group includes all F PNe, which are the dominant population in this galaxy \citep{Stanghellini2007}.} \citep{Stanghellini2007}, that MD and ORD PNe in our Galaxy are much more abundant near the Galactic centre, and that the highest metallicities and the highest abundances of nitrogen and helium are found in ORD and, especially, in MD PNe \citep{GarciaHernandez2014, DelgadoInglada2015}. Hence, our interpretation of the results presented in Fig.~\ref{fig:Sd_silicates} is that most ORD and MD PNe had high-metallicity progenitors, the most massive of which led to the formation of MD PNe. The high metallicity and/or mass of these MD PN progenitors leads to conditions where silicates can easily crystallize. The high-metallicity ORD and MD PNe initially contain both amorphous and crystalline silicates, but can evolve to form ORD and MD PNe that contain only crystalline silicates. Finally, newly-formed MD PNe have both amorphous and crystalline silicates, high surface brightness and small diameters.

\section{Conclusions}\label{sec:conclusion}

In this work, we have compiled the infrared spectrum of 267 PNe from the {\sl Spitzer}, {\sl ISO}, and {\sl IRAS} space telescopes. We have used these spectra to identify different dust features: PAHs, SiC, amorphous and crystalline silicates, and a broad feature around 30 \micron. With these identifications, we can classify 209 of the sample PNe into five groups: MD (PNe with PAHs and silicates), ORD (PNe with silicates), CRD (PNe with SiC and/or the 30 \micron\ feature, some of them with PAHs), oPAH (PNe with only PAHs) and F (featureless PNe).

Most of the objects in our sample are MD PNe (35 per cent), followed by ORD PNe (25 per cent), CRD and F PNe (17 per cent each), and oPAH PNe (6 per cent). Our sample is almost 40 per cent larger than samples analysed in previous works, and we compare the distributions of surface brightness and diameters of the different dust types as well as their distribution in the Galaxy using K-S tests to assess the significance of the similitudes and differences.

The distributions of nebular diameters and H$\alpha$ surface brightness indicate that oPAH and F PNe are significantly more evolved than the other groups, with F PNe being the most evolved objects, i.e., the ones with the lowest values of surface brightness and the largest diameters. Since oPAH and CRD PNe have similar distributions in the Galaxy, this suggests that oPAH PNe are evolved CRD PNe. On the other hand, and considering also the distribution in the Galaxy of the different dust types, F PNe could be evolved PNe from any initial dust type (though probably mainly evolved MD and ORD PNe). PNe with low-mass and low-metallicity progenitors, and PNe with limited dust formation because of a C/O ratio close to 1 in their ejecta, might also be F PNe, although we note that all F PNe show blackbody-like emission in their spectra. 

MD and ORD PNe are heavily concentrated towards the Galactic centre, and their distributions in the Galaxy are statistically very similar. CRD and oPAH PNe are more evenly distributed across the Galactic disc, and their distributions are also statistically similar to each other. We conclude that each pair of groups had very similar progenitors. We can also conclude that the formation of ORD and MD PNe is favoured by the higher metallicities near the Galactic centre, whereas lower metallicities favour the formation of CRD and oPAH PNe, in agreement with the predictions of models of stellar evolution, like those by \citet{2014MNRAS.445..347K}. 

We have also compared the distributions of MD and ORD PNe that have (i) only crystalline silicates in their spectra, (ii) both crystalline and amorphous silicates, and (iii) only amorphous silicates. There are no MD PNe with only amorphous silicates in their spectra, and the few ORD PNe with only amorphous silicates tend to be, on average, farther away from the Galactic centre and from the Galactic plane than the other MD and ORD PNe. This suggests that their progenitors were low-mass, low-metallicity objects. MD PNe with both crystalline and amorphous silicates have larger surface brightness and smaller diameters than all the other MD and ORD PNe. We suggest that they had the most massive and metal-rich progenitors, and that they can evolve to form MD PNe with only crystalline silicates in their spectra. ORD PNe have in general slightly lower surface brightness and slightly larger diameters than MD PNe, maybe indicating that their progenitors were not as massive or metal rich. ORD PNe with both crystalline and amorphous silicates might also evolve and form ORD PNe with only crystalline silicates, but the differences in the distributions of nebular diameters and H$\alpha$ surface brightness are not statistically significant in this case.

\section*{Acknowledgements}

We thank the anonymous referee for comments that helped to improve the paper. This work is based in part on observations made with the {\sl Spitzer Space Telescope}, which was operated by the Jet Propulsion Laboratory, California Institute of Technology under a contract with NASA. This research has made use of the NASA/IPAC Infrared Science Archive, which is funded by the National Aeronautics and Space Administration and operated by the California Institute of Technology. This work received partial support from Dirección General de Asuntos del Personal Académico-Programa de Apoyo a Proyectos de Investigación e Innovación Tecnológica (DGAPA-PAPIIT) project IN\,111423. D.H.-J. acknowledges support from a Mexican CONAHCYT scholarship.

\section*{Data Availability}
The data underlying this article will be shared on reasonable request to the corresponding author.



\bibliographystyle{mnras}
\bibliography{references} 


\appendix
\onecolumn
\onecolumn
\section{Spectra not previously published}\label{app:spectra}
In this appendix, we present PN spectra from the {\sl Spitzer Space Telescope} that have not been previously published. We do not show the data for four PNe with very noisy spectra (labelled as "poor quality" in Table~\ref{tab:dust_type}).

	\begin{figure}
		\center
		\includegraphics[width=\columnwidth]{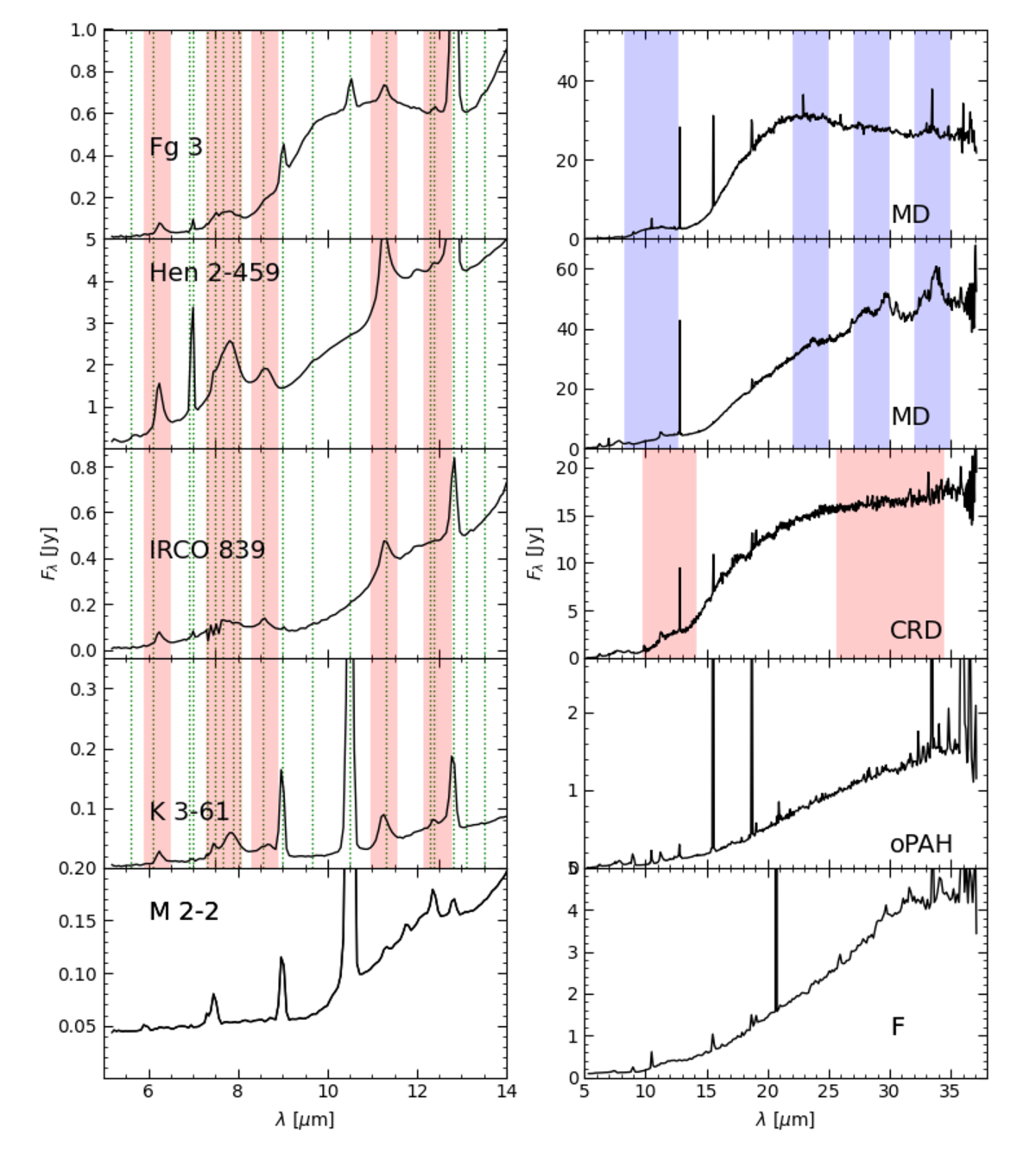}
    	\caption{Previously unpublished PN spectra from the {\sl Spitzer Space Telescope}. The left panels show the spectral region between 5 and 14 \micron, with the pink bands marking the positions of the PAH features and the green dotted lines the wavelengths of nebular lines (listed in Fig.~\ref{fig:PAH}). The right panels show the region from 5 to 38 \micron, with the pink bands around 12 and 30 \micron\ indicating the positions of the SiC and 30 \micron\ features, and the blue bands indicating the positions of the silicate features.}
   	 \label{fig:Espectra_A}
	\end{figure}

	\begin{figure}
		\center
		\includegraphics[width=\columnwidth]{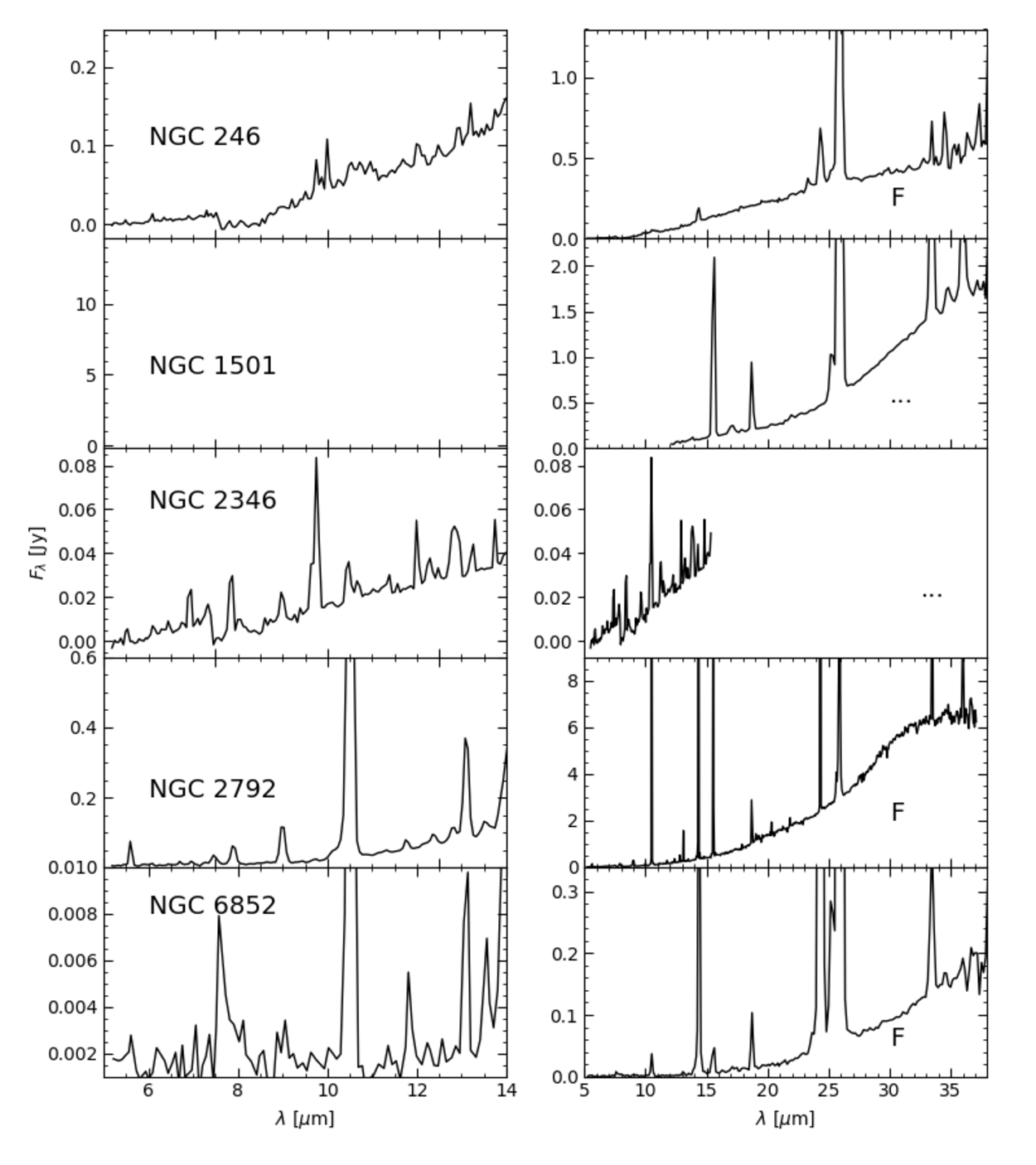}
    	\caption{Previously unpublished PN spectra from the {\sl Spitzer Space Telescope}. The left panels show the spectral region between 5 and 14 \micron. The right panels show the region from 5 to 38 \micron.}
   	 \label{fig:Espectra_B}
	\end{figure}

    \begin{figure}
		\center
		\includegraphics[width=\columnwidth]{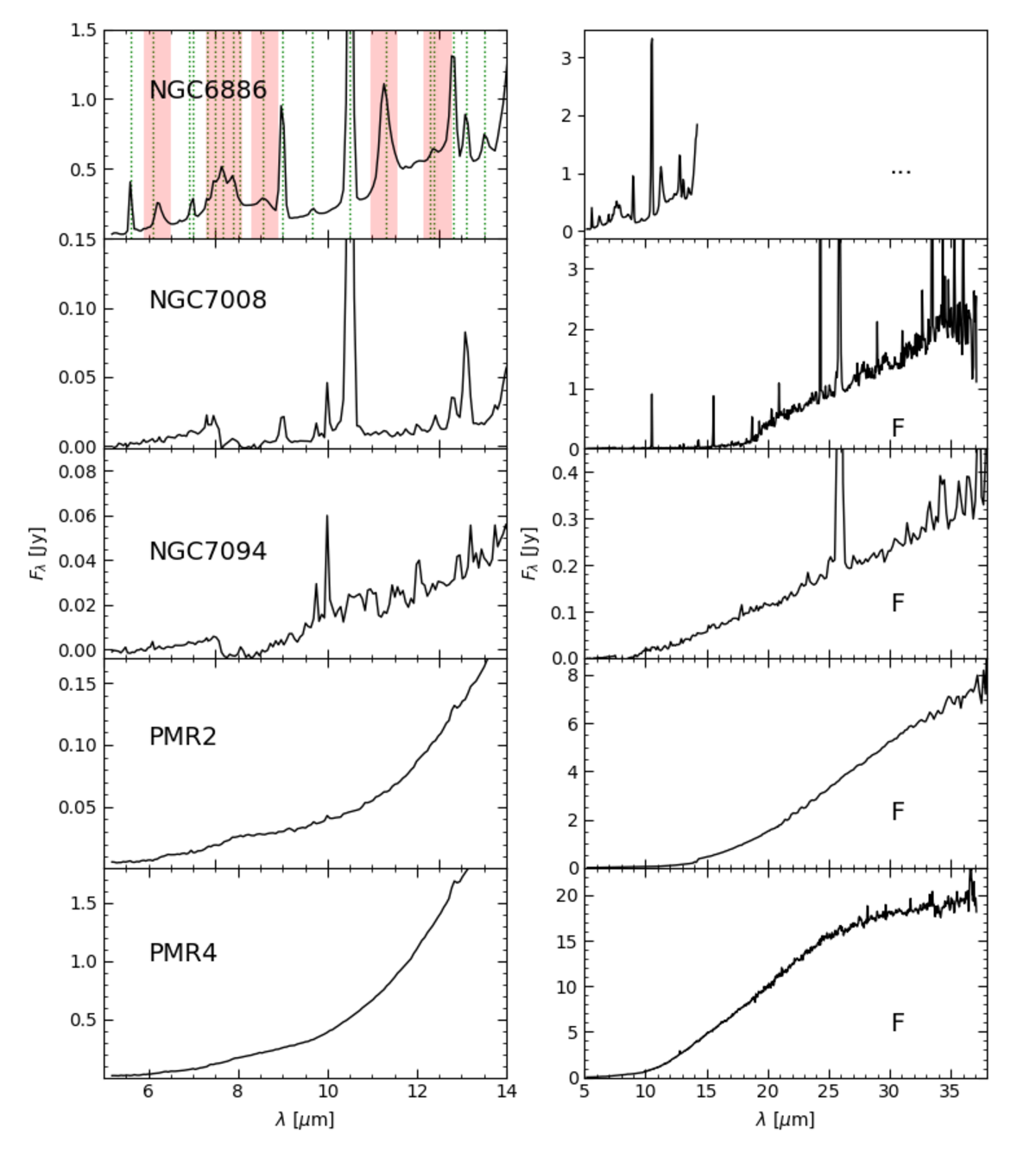}
    	\caption{Previously unpublished PN spectra from the {\sl Spitzer Space Telescope}. The left panels show the spectral region between 5 and 14 \micron, with the pink bands marking the positions of the PAH features and the green dotted lines the wavelengths of nebular lines (listed in Fig.~\ref{fig:PAH}). The right panels show the region from 5 to 38 \micron.}
   	 \label{fig:Espectra_D}
	\end{figure}

\clearpage
\section{Dust types}\label{app:dust}

In this appendix, we present the identifications of the dust features for each PN and its classification according to these features. In the first and second columns of the table, we identify the PNe with their common name and PN\,G number. In the third column, we indicate the dust type (MD, ORD, CRD, PAH, or F) if it could be determined. In columns 4 to 8, the feature identifications of PAHs, SiC, 30 \micron, and crystalline and amorphous silicates are indicated. The last two columns list the telescope and the reference for the published spectrum, respectively.

\begin{longtable}{l l c c c c c c l l}
\caption{Dust features and final classification of the PNe.}
\label{tab:dust_type}\\ 
\toprule
Name & PN\,G & Type & PAH & SiC & 30 $\mu$m & Am. & Crys.  & Origin & Ref. \\
          &   &      & &     &           & sil.   & sil.   & &\\
\hline \hline
\endfirsthead
\multicolumn{10}{c}{Table B1 (continued): Dust features and final classification of the PNe.} \\ 
\hline
Name & PN\,G & Type & PAH & SiC & 30 $\mu$m & Am. & Crys.  & Origin & Ref. \\
          &   &      & &     &           & sil.   & sil.   & &\\
\hline \hline
\endhead

\hline
\endfoot
\endlastfoot

19W32  &  359.2$+$01.2  & F?  &  \ding{53}  &  \ding{53}  &  \ding{53}?  &  \checkmark  &  \ding{53}? &  \small{Spitzer}  &  PC09 	\\
Abell\,43  &  036.0$+$17.6  & $\cdots$  &  $\cdots$  &  \ding{53}  &  \ding{53}  &  \ding{53}  &  \ding{53} &  \small{Spitzer}  &  poor quality 	\\
Al\,2$-$E  &  359.3$+$03.6  &  ORD  &  \ding{53}  &  \ding{53}  &  \ding{53}  &  \ding{53}  &  \checkmark &  \small{Spitzer}  &  S12  	\\
Al\,2$-$F  &  358.5$+$02.9  &  F  &  \ding{53}  &  \ding{53}  &  \ding{53}  &  \ding{53}  &  \ding{53} &  \small{Spitzer}  &  S12 	\\
Al\,2$-$R  &  358.7$-$02.7  &  ORD  &  \ding{53}  &  \ding{53}  &  \ding{53}  &  \checkmark  &  \ding{53} &  \small{Spitzer}  &  S12 	\\
Bl\,3$-$15  &  000.6$-$01.3  &  oPAH  &  \checkmark  &  \ding{53}  &  \ding{53}  &  \ding{53}  &  \ding{53} &  \small{Spitzer}  &  S12 	\\
Bl\,B  &  358.3$+$01.2  &  MD  &  \checkmark  &  \ding{53}  &  \ding{53}  &  \ding{53}  &  \checkmark  &  \small{Spitzer}  &  S12	\\
Bl\,Q  &  001.6$-$01.3  &  ORD  &  \ding{53}  &  \ding{53}  &  \ding{53}  &  \ding{53}  &  \checkmark &  \small{Spitzer}  &  G08  	\\
BoBn\,1  &  108.4$-$76.1  &  oPAH  &  \checkmark  &  \ding{53}  &  \ding{53}  &  \ding{53}  &  \ding{53} &  \small{Spitzer}  &  S12  	\\
Cn\,1$-$1  &  330.7$+$04.1  & $\cdots$ &  \ding{53}? &  \ding{53}? &  $\cdots$  &  \ding{53}  &  $\cdots$ &  \small{IRAS}  & P84 	\\
Cn\,1$-$3  &  345.0$-$04.9  &  ORD  &  \ding{53}  &  \ding{53}  &  \ding{53}  &  \ding{53}  &  \checkmark &  \small{Spitzer}  &  PC09 	\\
Cn\,1$-$5  &  002.2$-$09.4  &  MD  &  \checkmark  &  \ding{53}  &  \ding{53}  &  \ding{53}  &  \checkmark &  \small{Spitzer}  &  PC09 	\\
Cn\,2$-$1  &  356.2$-$04.4  &  ORD  &  \ding{53}  &  \ding{53}  &  \ding{53}  &  \checkmark  &  \checkmark &  \small{Spitzer}  &  S12 	\\
CRL\,618  &  166.4$-$06.5  & $\cdots$  &  \ding{53}  &  \ding{53}  &  $\cdots$  &  \ding{53}  &  $\cdots$ &  \small{IRAS}  & P86 	\\
CTS\,1  &  019.8$+$05.6  &  F  &  \ding{53}  &  \ding{53}  &  \ding{53}  &  \ding{53}  &  \ding{53} &  \small{Spitzer}  &  S12 	\\
DdDm\,1  &  061.9$+$41.3  &  ORD  & \ding{53} &  \ding{53}&  \ding{53}  &  \checkmark    &  \checkmark    &  \small{Spitzer}  &  DI14 	\\
Fg\,3  &  352.9$-$07.5  &  MD  &  \checkmark  &  \ding{53}  &  \ding{53}  &  \checkmark  &  \checkmark &  \small{Spitzer}  &  t.w.  	\\
GJJC\,1  &  009.8$-$07.5  &  F  &  \ding{53}  &  \ding{53}  &  \ding{53}  &  \ding{53}  &  \ding{53} &  \small{Spitzer}  &  S12  	\\
H\,1$-$4  &  351.3$+$07.6  &  ORD  &  \ding{53}  &  \ding{53}  &  \ding{53}  &  \checkmark  &  \ding{53} &  \small{Spitzer}  &  S12 	\\
H\,1$-$8  &  352.6$+$03.0  &  MD  &  \checkmark  &  \ding{53}  &  \ding{53}  &  \ding{53}  &  \checkmark &  \small{Spitzer}  &  S12 	\\
H\,1$-$9  &  355.9$+$03.6  &  ORD  &  \ding{53}  &  \ding{53}  &  \ding{53}  &  \checkmark  &  \ding{53} &  \small{Spitzer}  &  S12  	\\
H\,1$-$15  &  001.4$+$05.3  &  ORD  &  \ding{53}  &  \ding{53}  &  \ding{53}  &  \ding{53}  &  \checkmark &  \small{Spitzer}  &  G08 	\\
H\,1$-$16  &  000.1$+$04.3  &  MD  &  \checkmark  &  \ding{53}  &  \ding{53}  &  \ding{53}  &  \checkmark &  \small{Spitzer}  &  PC09 	\\
H\,1$-$17  &  358.3$+$01.2  &  MD  &  \checkmark  &  \checkmark  &  \ding{53}  &  \checkmark  &  \checkmark &  \small{Spitzer}  &  S12  	\\
H\,1$-$18  &  357.6$+$02.6  &  MD  &  \checkmark  &  \ding{53}  &  \ding{53}  &  \ding{53}  &  \checkmark &  \small{Spitzer}  &  S12 	\\
H\,1$-$19  &  358.9$+$03.4  &  MD  &  \checkmark  &  \ding{53}  &  \ding{53}  &  \ding{53}  &  \checkmark &  \small{Spitzer}  &  S12 	\\
H\,1$-$20  &  358.9$+$03.2  &  MD  &  \checkmark  &  \ding{53}  &  \ding{53}  &  \ding{53}  &  \checkmark &  \small{Spitzer}  &  G08 	\\
H\,1$-$21  &  348.4$-$04.1  &  MD  &  \checkmark  &  \ding{53}  &  \ding{53}  &  \ding{53}  &  \checkmark &  \small{Spitzer}  &  S12 	\\
H\,1$-$22  &  350.8$-$02.4  &  ORD  &  \ding{53}  &  \ding{53}  &  \ding{53}  &  \ding{53}  &  \checkmark &  \small{Spitzer}  &  S12 	\\
H\,1$-$23  &  357.6$+$01.7  &  ORD  &  \ding{53}  &  \ding{53}  &  \ding{53}  &  \checkmark  &  \checkmark &  \small{Spitzer}  &  S12 	\\
H\,1$-$29  &  355.2$-$02.5  &  MD  &  \checkmark  &  \ding{53}  &  \ding{53}  &  \ding{53}  &  \checkmark &  \small{Spitzer}  &  S12 	\\
H\,1$-$32  &  355.6$-$02.7  &  ORD  &  \ding{53}  &  \ding{53}  &  \ding{53}  &  \checkmark  &  \checkmark &  \small{Spitzer}  &  PC09 	\\
H\,1$-$33  &  355.7$-$03.0  &  MD  &  \checkmark  &  \ding{53}  &  \ding{53}  &  \ding{53}  &  \checkmark &  \small{Spitzer}  &  S12  	\\
H\,1$-$34  &  005.5$+$02.7  &  MD  &  \checkmark  &  \ding{53}  &  \ding{53}  &  \ding{53}  &  \checkmark &  \small{Spitzer}  &  S12 	\\
H\,1$-$35  &  355.7$-$03.5  &  ORD  &  \ding{53}  &  \ding{53}  &  \ding{53}  &  \checkmark  &  \ding{53} &  \small{Spitzer}  &  PC09 	\\
H\,1$-$40  &  359.7$-$02.6  &  MD  &  \checkmark  &  \ding{53}  &  \ding{53}  &  \checkmark  &  \checkmark &  \small{Spitzer}  &  PC09 	\\
H\,1$-$43  &  357.1$-$04.7  &  MD  &  \checkmark  &  \ding{53}  &  \ding{53}  &  \ding{53}  &  \checkmark &  \small{Spitzer}  &  PC09 	\\
H\,1$-$44  &  358.9$-$03.7  &  MD  &  \checkmark  &  \ding{53}  &  \ding{53}  &  \ding{53}  &  \checkmark &  \small{Spitzer}  &  S12 	\\
H\,1$-$46  &  358.5$-$04.2  &  ORD  &  \ding{53}  &  \ding{53}  &  \ding{53}  &  \checkmark  &  \checkmark &  \small{Spitzer}  &  S12 	\\
H\,1$-$53  &  004.3$-$02.6  &  MD  &  \checkmark  &  \ding{53}  &  \ding{53}  &  \ding{53}  &  \checkmark &  \small{Spitzer}  &  S12 	\\
H\,1$-$56  &  001.7$-$04.6  &  ORD  &  \ding{53}  &  \ding{53}  &  \ding{53}  &  \ding{53}  &  \checkmark &  \small{Spitzer}  &  S12 	\\
H\,1$-$61  &  006.5$-$03.1  &  MD  &  \checkmark  &  \ding{53}  &  \ding{53}  &  \ding{53}  &  \checkmark &  \small{Spitzer}  &  PC09 	\\
H\,2$-$11  &  000.7$+$04.7  &  MD  &  \checkmark  &  \ding{53}  &  \ding{53}  &  \ding{53}  &  \checkmark &  \small{Spitzer}  &  G08 	\\
H\,2$-$13  &  357.2$+$02.0  &  MD  &  \checkmark  &  \ding{53}  &  \ding{53}  &  \ding{53}  &  \checkmark &  \small{Spitzer}  &  S12 	\\
H\,2$-$17  &  003.1$+$03.4  &  MD  &  \checkmark  &  \ding{53}  &  \ding{53}  &  \ding{53}  &  \checkmark &  \small{Spitzer}  &  S12 	\\
H\,2$-$27  &  356.5$-$03.6  &  ORD  &  \ding{53}  &  \ding{53}  &  \ding{53}  &  \ding{53}  &  \checkmark &  \small{Spitzer}  &  S12 	\\
H\,2$-$31  &  001.7$-$01.6  &  ORD  &  \ding{53}  &  \ding{53}  &  \checkmark  &  \ding{53}  &  \ding{53} &  \small{Spitzer}  &  S12 	\\
H\,2$-$32  &  000.6$-$02.3  &  MD  &  \checkmark  &  \ding{53}  &  \ding{53}  &  \checkmark  &  \checkmark &  \small{Spitzer}  &  S12 	\\
H\,2$-$39  &  002.9$-$03.9  &  F  &  \ding{53}  &  \ding{53}  &  \ding{53}  &  \ding{53}  &  \ding{53} &  \small{Spitzer}  &  S12  	\\
H\,2$-$46  &  000.8$-$07.6  &  oPAH  &  \checkmark  &  \ding{53}  &  \ding{53}  &  \ding{53}  &  \ding{53} &  \small{Spitzer}  &  S12 	\\
H\,2$-$48  &  011.3$-$09.4  &  ORD  &  \ding{53}  &  \ding{53}  &  \ding{53}  &  \checkmark  &  \ding{53} &  \small{Spitzer}  &  S12 	\\
Hb\,4  &  003.1$+$02.9  &  ORD  &  \ding{53}  &  \ding{53}  &  \ding{53}  &  \ding{53}  &  \checkmark &  \small{Spitzer}  &  PC09 	\\
Hb\,5  &  359.3$-$00.9  &  CRD  &  $\cdots$  &  \checkmark  &  \checkmark  &  \ding{53}  &  \ding{53} &  \small{ISO}  & H02a 	\\
Hb\,6  &  007.2$+$01.8  &  ORD  &  \ding{53}  &  \ding{53}  &  \ding{53}  &  \ding{53}  &  \checkmark &  \small{Spitzer}  &  PC09 	\\
Hb\,7  &  003.9$-$14.9  &  ORD  &  \ding{53}  &  \ding{53}  &  \ding{53}  &  \checkmark  &  \ding{53} &  \small{Spitzer}  &  S12 	\\
Hb\,12  &  111.8$-$02.8  & ORD? &  \ding{53}?  &  \ding{53}  &  $\cdots$  &  \checkmark  &  $\cdots$ &  \small{IRAS}  & P86 	\\
Hen\,2$-$5  &  264.4$-$12.7  &  CRD  &  \ding{53}  &  \ding{53}  &  \checkmark  &  \ding{53}  &  \ding{53} &  \small{Spitzer}  &  S12 	\\
Hen\,2$-$21  &  275.3$-$04.7  &  CRD  &  \ding{53}  &  \ding{53}  &  \checkmark  &  \ding{53}  &  \ding{53} &  \small{Spitzer}  &  S12 	\\
Hen\,2$-$62  &  295.3$-$09.3  &  ORD  &  \ding{53}  &  \ding{53}  &  \ding{53}  &  \checkmark  &  \ding{53} &  \small{Spitzer}  &  S12 	\\
Hen\,2$-$63  &  289.8$+$07.7  &  oPAH  &  \checkmark  &  \ding{53}  &  \ding{53}  &  \ding{53}  &  \ding{53} &  \small{Spitzer}  &  S12  	\\
Hen\,2$-$68 &  294.9$-$04.3  &  CRD  &  \checkmark  &  \checkmark  &  \checkmark  &  \ding{53}  &  \ding{53} &  \small{Spitzer}  &  S12 	\\
Hen\,2$-$73  &  296.3$-$03.0  &  MD  &  \checkmark  &  \ding{53}  &  \ding{53}  &  \checkmark  &  \checkmark &  \small{Spitzer}  &  S12 	\\
Hen\,2$-$78  &  297.4$+$03.7  &  oPAH  &  \checkmark  &  \ding{53}  &  \ding{53}  &  \ding{53}  &  \ding{53} &  \small{Spitzer}  &  S12 	\\
Hen\,2$-$86  &  300.7$-$02.0  &  MD  &  \checkmark  &  \ding{53}  &  \ding{53}  &  \checkmark  &  \checkmark &  \small{Spitzer}  &  S12 	\\
Hen\,2$-$95  &  307.5$-$04.9  &  MD  &  \checkmark  &  \ding{53}  &  \ding{53}  &  \checkmark  &  \checkmark &  \small{Spitzer}  &  S12 	\\
Hen\,2$-$99  &  309.0$-$04.2  &  oPAH  &  \checkmark  &  \ding{53}  &  \ding{53}  &  \ding{53}  &  \ding{53} &  \small{Spitzer}  &  S12  	\\
Hen\,2$-$108  &  316.1$+$08.4  &  ORD  &  \ding{53}  &  \ding{53}  &  \ding{53}  &  \checkmark  &  \checkmark &  \small{Spitzer}  &  S12 	\\
Hen\,2$-$113  &  321.0$+$03.9  &  MD? &  \checkmark?  &  \ding{53}  &  \ding{53}  &  \checkmark  &  \checkmark &  \small{ISO/IRAS}  & W98, M99 	\\
Hen\,2$-$115  &  321.3$+$02.8  &  CRD  &  \ding{53}  &  \checkmark  &  \checkmark  &  \ding{53}  &  \ding{53} &  \small{Spitzer}  &  S12 	\\
Hen\,2$-$125  &  324.2$+$02.5  &  MD  &  \checkmark  &  \ding{53}  &  \ding{53}  &  \ding{53}  &  \checkmark &  \small{Spitzer}  &  S12 	\\
Hen\,2$-$129  &  325.0$+$03.2  &  ORD  &  \ding{53}  &  \ding{53}  &  \ding{53}  &  \ding{53}  &  \checkmark &  \small{Spitzer}  &  S12 	\\
Hen\,2$-$133  &  324.8$-$01.1  &  MD  &  \checkmark  &  \ding{53}  &  \ding{53}  &  \checkmark  &  \checkmark &  \small{Spitzer}  &  S12 	\\
Hen\,2$-$140  &  327.1$-$01.8  &  MD  &  \checkmark  &  \ding{53}  &  \ding{53}  &  \ding{53}  &  \checkmark &  \small{Spitzer}  &  S12 	\\
Hen\,2$-$149  &  329.4$-$02.7  &  F  &  \ding{53}  &  \ding{53}  &  \ding{53}  &  \ding{53}  &  \ding{53} &  \small{Spitzer}  &  S12 	\\
Hen\,2$-$151  &  326.0$-$06.5  &  ORD  &  \ding{53}  &  \ding{53}  &  \ding{53}  &  \checkmark  &  \checkmark &  \small{Spitzer}  &  S12 	\\
Hen\,2$-$157  &  331.0$-$02.7  &  ORD  &  \ding{53}  &  \ding{53}  &  \ding{53}  &  \checkmark  &  \checkmark &  \small{Spitzer}  &  S12 	\\
Hen\,2$-$178  &  344.2$+$04.7  &  ORD?  &  \checkmark?  &  \ding{53}  &  \ding{53}  &  \checkmark  &   \ding{53}  &  \small{Spitzer}  &  S12	\\
Hen\,2$-$186  &  336.3$-$05.6  &  MD  &  \checkmark  &  \ding{53}  &  \ding{53}  &  \ding{53}  &  \checkmark &  \small{Spitzer}  &  S12  	\\
Hen\,2$-$248  &  341.5$-$09.1  &  F  &  \ding{53}  &  \ding{53}  &  \ding{53}  &  \ding{53}  &  \ding{53} &  \small{Spitzer}  &  S12 	\\
Hen\,2$-$26  &  278.6$-$06.7  &  CRD  &  \checkmark?  &  \checkmark  &  \checkmark  &  \ding{53}  &  \ding{53} &  \small{Spitzer}  &  S12 	\\
Hen\,2$-$260  &  008.2$+$06.8  &  ORD  &  \ding{53}  &  \ding{53}  &  \ding{53}  &  \checkmark  &  \ding{53} &  \small{Spitzer}  &  PC09 	\\
Hen\,2$-$262  &  001.2$+$02.1  &  ORD  &  \ding{53}  &  \ding{53}  &  \ding{53}  &  \ding{53}  &  \checkmark &  \small{Spitzer}  &  G08 	\\
Hen\,2$-$306  &  348.8$-$09.0  &  F  &  \ding{53}  &  \ding{53}  &  \ding{53}  &  \ding{53}  &  \ding{53} &  \small{Spitzer}  &  S12 	\\
Hen\,2$-$406  &  008.6$-$07.0  &  F  &  \ding{53}  &  \ding{53}  &  \ding{53}  &  \ding{53}  &  \ding{53} &  \small{Spitzer}  &  S12 	\\
Hen\,2$-$438   &  064.7$+$05.0  &  MD  &  \checkmark  &  \ding{53}  &  \ding{53}  &  \checkmark  &  \checkmark &  \small{ISO}  & BS05 	\\
Hen\,2$-$440  &  060.5$+$01.8  &  MD  &  \checkmark  &  \ding{53}  &  \ding{53}  &  \checkmark  &  \checkmark &  \small{Spitzer}  &  S12  	\\
Hen\,2$-$459  &  068.3$-$02.7  &  MD  &  \checkmark  &  \ding{53}  &  \ding{53}  &  \ding{53}  &  \checkmark &  \small{Spitzer}  &  t.w. 	\\
Hen\,3$-$1132  &  336.9$+$08.3  &  ORD?  &  \ding{53}  &  \ding{53}  &  \ding{53}  &  \checkmark?  &  \ding{53} &  \small{Spitzer}  &  S12 	\\
Hen\,3$-$1333  &  332.9$-$09.9  & $\cdots$  &  \ding{53}  &  \ding{53}  &  $\cdots$  &  \ding{53}  &  $\cdots$ &  \small{IRAS}  & P86	\\
Hen\,3$-$1716  &  014.0$-$05.5  &  F  &  \ding{53}  &  \ding{53}  &  \ding{53}  &  \ding{53}  &  \ding{53} &  \small{Spitzer}  &  S12 	\\
Hu\,2$-$1  &  051.4$+$09.6  &  CRD  &  \ding{53}  &  \checkmark  &  $\cdots$  &  \ding{53}  &  $\cdots$ &  \small{Spitzer}  &  DI14 	\\
IC\,418  &  215.2$-$24.2  &  CRD  &  \ding{53}  &  \checkmark  &  \ding{53}  &  \ding{53}  &  \ding{53} &  \small{Spitzer/ISO}  & H02a, DI14 	\\
IC\,2003  &  161.2$-$14.8  &  MD  &  \checkmark  &  \ding{53}  &  \ding{53}  &  \ding{53}  &  \checkmark &  \small{Spitzer}  &  S12 	\\
IC\,4732  &  010.7$-$06.4  &  MD  &  \checkmark?  &  \ding{53}  &  \ding{53}  &  \checkmark  &  \checkmark &  \small{Spitzer}  &  PC09 	\\
IC\,4776  &  002.0$-$13.4  &  MD  &  \checkmark  &  \ding{53}  &  \ding{53}  &  \checkmark  &  \checkmark &  \small{Spitzer}  &  PC09  	\\
IC\,4846  &  027.6$-$09.6  &  ORD  &  \ding{53}  &  \ding{53}  &  \ding{53}  &  \checkmark?  &  \ding{53} &  \small{Spitzer}  &  S12 	\\
IC\,4997  &  058.3$-$10.9  & $\cdots$  &  \ding{53}?  &  \ding{53}?  &  $\cdots$  &  \ding{53}?  &  $\cdots$ &  \small{IRAS}  & B83 	\\
IC\,5117  &  089.8$-$05.1  & $\cdots$ & $\cdots$ &  \ding{53}  & $\cdots$ &  \ding{53}  &  $\cdots$ &  \small{IRAS}  & Z90 	\\
IRCO\,839  &  110.1$+$01.9  &  CRD  &  \checkmark  &  \checkmark  &  \ding{53}  &  \ding{53}  &  \ding{53} &  \small{Spitzer}  &  t.w. 	\\
K\,2$-$16  &  352.9$+$11.4  & $\cdots$  &  $\cdots$  &  \ding{53}  &  \checkmark  &  \ding{53}  &  \ding{53} &  \small{ISO}  & H02a 	\\
K\,3$-$6  &  031.0$+$04.1  &  MD  &  \checkmark  &  \ding{53}  &  \ding{53}  &  \checkmark  &  \checkmark &  \small{Spitzer}  &  S12 	\\
K\,3$-$11  &  023.8$-$01.7  &  MD  &  \checkmark  &  \ding{53}  &  \ding{53}  &  \ding{53}  &  \checkmark &  \small{Spitzer}  &  S12 	\\
K\,3$-$13  &  034.0$+$02.2  &  ORD  &  \ding{53}  &  \ding{53}  &  \ding{53}  &  \ding{53}  &  \checkmark &  \small{Spitzer}  &  S12 	\\
K\,3$-$14  &  042.0$+$05.4  &  ORD  &  \ding{53}  &  \ding{53}  &  \ding{53}  &  \checkmark  &  \ding{53} &  \small{Spitzer}  &  S12 	\\
K\,3$-$15  &  041.8$+$04.4  &  CRD  &  \checkmark  &  \checkmark  &  \ding{53}  &  \ding{53}  &  \ding{53} &  \small{Spitzer}  &  S12  	\\
K\,3$-$17  &  039.8$+$02.1  &  MD  &  \checkmark  &  \ding{53}  &  \ding{53}  &  \ding{53}  &  \checkmark &  \small{Spitzer/ISO}  &  S12  	\\
K\,3$-$19  &  032.9$-$02.8  &  CRD  &  \checkmark  &  \checkmark  &  \checkmark  &  \ding{53}  &  \ding{53} &  \small{Spitzer}  &  S12 	\\
K\,3$-$20  &  032.5$-$03.2  &  MD  &  \checkmark  &  \ding{53}  &  \ding{53}  &  \ding{53}  &  \checkmark &  \small{Spitzer}  &  S12 	\\
K\,3$-$29  &  048.1$+$01.1  &  CRD  &  \checkmark  &  \ding{53}  &  \checkmark  &  \ding{53}  &  \ding{53} &  \small{Spitzer}  &  S12 	\\
K\,3$-$31  &  052.9$+$02.7  &  CRD  &  \checkmark  &  \checkmark  &  \checkmark  &  \ding{53}  &  \ding{53} &  \small{Spitzer}  &  S12 	\\
K\,3$-$33  &  045.9$-$01.9  &  MD  &  \checkmark  &  \ding{53}  &  \ding{53}  &  \ding{53}  &  \checkmark &  \small{Spitzer}  &  S12 	\\
K\,3$-$37  &  059.4$+$02.3  &  CRD  &  \ding{53}  &  \checkmark  &  \checkmark  &  \ding{53}  &  \ding{53} &  \small{Spitzer}  &  S12 	\\
K\,3$-$39  &  059.9$+$02.0  &  CRD  &  \checkmark  &  \checkmark  &  \checkmark  &  \ding{53}  &  \ding{53} &  \small{Spitzer}  &  S12 	\\
K\,3$-$40  &  058.9$+$01.3  &  ORD  &  \ding{53}  &  \ding{53}  &  \ding{53}  &  \ding{53}  &  \checkmark &  \small{Spitzer}  &  S12 	\\
K\,3$-$43  &  055.1$-$01.8  &  F  &  \ding{53}  &  \ding{53}  &  \ding{53}  &  \ding{53}  &  \ding{53} &  \small{Spitzer}  &  S12 	\\
K\,3$-$49  &  069.2$+$02.8  &  ORD  &  \ding{53}  &  \ding{53}  &  \ding{53}  &  \checkmark  &  \checkmark &  \small{Spitzer}  &  S12 	\\
K\,3$-$52  &  067.9$-$00.2  &  MD  &  \checkmark  &  \ding{53}  &  \ding{53}  &  \ding{53}  &  \checkmark &  \small{Spitzer}  &  S12 	\\
K\,3$-$54  &  063.8$-$03.3  &  CRD  &  \checkmark?  &  \checkmark  &  \checkmark  &  \ding{53}  &  \ding{53} &  \small{Spitzer}  &  S12 	\\
K\,3$-$56  &  079.9$+$06.4  &  CRD  &  \ding{53}  &  \ding{53}?  &  \checkmark  &  \ding{53}  &  \ding{53} &  \small{Spitzer}  &  S12 	\\
K\,3$-$60  &  098.2$+$04.9  &  CRD  &  \checkmark  &  \checkmark  &  \checkmark  &  \ding{53}  &  \ding{53} &  \small{Spitzer}  &  S12  	\\
K\,3$-$61  &  096.3$+$02.3  &  oPAH  &  \checkmark  &  \ding{53}  &  \ding{53}  &  \ding{53}  &  \ding{53} &  \small{Spitzer}  &  t.w. 	\\
K\,3$-$62  &  095.2$+$00.7  &  CRD  &  \checkmark  &  \checkmark  &  \checkmark  &  \ding{53}  &  \ding{53} &  \small{Spitzer}  &  S12 	\\
K\,3$-$78  &  088.7$+$04.6  &  F  &  \ding{53}  &  \ding{53}  &  \ding{53}  &  \ding{53}  &  \ding{53} &  \small{Spitzer}  &  S12 	\\
K\,3$-$87  &  107.4$-$02.6  &  CRD?  &  \ding{53}  &  \ding{53}  &  \checkmark?  &  \ding{53}  &  \ding{53} &  \small{Spitzer}  &  S12 	\\
K\,4$-$8  &  025.3$-$04.6  &  F  &  \ding{53}  &  \ding{53}  &  \ding{53}  &  \ding{53}  &  \ding{53} &  \small{Spitzer}  &  S12 	\\
K\,4$-$16  &  048.5$+$04.2  &  CRD?  &  \ding{53}  &  \ding{53}  &  \checkmark?  &  \ding{53}  &  \ding{53} &  \small{Spitzer}  &  S12 	\\
K\,4$-$19  &  038.4$-$03.3  &  F  &  \ding{53}  &  \ding{53}  &  \ding{53}  &  \ding{53}  &  \ding{53} &  \small{Spitzer}  &  S12 	\\
K\,4$-$41  &  068.7$+$01.9  &  ORD  &  \ding{53}  &  \ding{53}  &  \ding{53}  &  \ding{53}  &  \checkmark &  \small{Spitzer}  &  S12 	\\
K\,5$-$4  &  351.9$-$01.9  &  MD  &  \checkmark  &  \ding{53}  &  \ding{53}  &  \checkmark  &  \checkmark &  \small{Spitzer}  &  S12  	\\
KFL\,4  &  003.0$-$02.6  &  CRD  &  \ding{53}  &  \ding{53}  &  \checkmark  &  \ding{53}  &  \ding{53} &  \small{Spitzer}  &  S12 	\\
KFL\,12  &  003.2$-$04.4  &  F  &  \ding{53}  &  \ding{53}  &  \ding{53}  &  \ding{53}  &  \ding{53} &  \small{Spitzer}  &  S12 	\\
KW97\,58$-$15  &  093.9$-$00.1  &  oPAH  &  \checkmark  &  \ding{53}  &  \ding{53}  &  \ding{53}  &  \ding{53} &  \small{ISO/IRAS}  & P86, W98 	\\
M\,1$-$5  &  184.0$-$02.1  &  CRD  &  \checkmark  &  \checkmark  &  \checkmark  &  \ding{53}  &  \ding{53} &  \small{Spitzer}  &  S12 	\\
M\,1$-$6  &  211.2$-$03.5  & ORD?  &  \ding{53}  &  \ding{53}  & $\cdots$  &  \checkmark &  $\cdots$ &  \small{IRAS}  & O14 	\\
M\,1$-$11  &  232.8$-$04.7  & $\cdots$ & $\cdots$ &  \ding{53} &  $\cdots$  &  $\cdots$  &  $\cdots$ &  \small{IRAS}  & Z90 	\\
M\,1$-$12  &  235.3$-$03.9  &  CRD  &  \ding{53}  &  \checkmark  &  \ding{53}  &  \ding{53}  &  \ding{53} &  \small{Spitzer}  &  S12 	\\
M\,1$-$20  &  006.1$+$08.3  &  CRD  &  \checkmark  &  \checkmark  &  \checkmark  &  \ding{53}  &  \ding{53} &  \small{Spitzer}  &  PC09 	\\
M\,1$-$25  &  004.9$+$04.9  &  MD  &  \checkmark  &  \ding{53}  &  \ding{53}  &  \ding{53}  &  \checkmark &  \small{Spitzer}  &  PC09 	\\
M\,1$-$26  &  358.9$-$00.7  & $\cdots$ &  \ding{53}?  &  \checkmark? &  $\cdots$  &  \ding{53}  &  $\cdots$  &  \small{IRAS}  & Z90 	\\
M\,1$-$27  &  356.5$-$02.3  &  MD  &  \checkmark  &  \ding{53}  &  \ding{53}  &  \ding{53}  &  \checkmark &  \small{Spitzer}  &  PC09 	\\
M\,1$-$31  &  006.4$+$02.0  &  MD  &  \checkmark  &  \ding{53}  &  \ding{53}  &  \ding{53}  &  \checkmark &  \small{Spitzer}  &  PC09 	\\
M\,1$-$32  &  011.9$+$04.2  & $\cdots$  &  \ding{53}  &  $\cdots$  &  $\cdots$  &  $\cdots$  &  $\cdots$ &  \small{Spitzer}  &  PC09 	\\
M\,1$-$33  &  013.1$+$04.1  &  ORD  &  \ding{53}  &  \ding{53}  &  \ding{53}  &  \checkmark?  &  \checkmark &  \small{Spitzer}  &  PC09 	\\
M\,1$-$35  &  107.4$-$06.6  & $\cdots$  &  $\cdots$  &  \ding{53}  &  \ding{53}  &  \ding{53}  &  \checkmark &  \small{Spitzer}  &  S12  	\\
M\,1$-$37  &  002.6$-$03.4  &  MD  &  \checkmark  &  \ding{53}  &  \ding{53}  &  \ding{53}  &  \checkmark &  \small{Spitzer}  &  S12 	\\
M\,1$-$38  &  002.4$-$03.7  &  MD  &  \checkmark  &  \ding{53}  &  \ding{53}  &  \ding{53}  &  \checkmark &  \small{Spitzer}  &  S12 	\\
M\,1$-$40  &  008.3$-$01.1  &  MD  &  \checkmark  &  \ding{53}  &  \ding{53}  &  \ding{53}  &  \checkmark &  \small{Spitzer}  &  PC09 	\\
M\,1$-$42  &  002.7$-$04.8  & MD  &  \checkmark  &  \ding{53}  &  \ding{53}  &  \ding{53}  &  \checkmark  &  \small{Spitzer}  & DI14 	\\
M\,1$-$45  &  012.6$-$02.7  &  MD  &  \checkmark  &  \ding{53}  &  \ding{53}  &  \ding{53}  &  \checkmark &  \small{Spitzer}  &  S12 	\\
M\,1$-$51  &  020.9$-$01.1  &  MD  &  \checkmark  &  \ding{53}  &  \ding{53}  &  \ding{53}  &  \checkmark &  \small{Spitzer}  &  PC09 	\\
M\,1$-$60  &  019.7$-$04.5  &  MD  &  \checkmark  &  \ding{53}  &  \ding{53}  &  \ding{53}  &  \checkmark &  \small{Spitzer}  &  PC09 	\\
M\,1$-$61  &  019.4$-$05.3  &  MD  &  \checkmark  &  \ding{53}  &  \ding{53}  &  \checkmark  &  \checkmark &  \small{Spitzer}  &  S12 	\\
M\,1$-$62  &  012.5$-$09.8  &  F?  &  \ding{53}  &  \ding{53}  &  \ding{53}?  &  \ding{53}  &  \ding{53} &  \small{Spitzer}  &  S12 	\\
M\,1$-$65  &  043.1$+$03.8  &  F  &  \ding{53}  &  \ding{53}  &  \ding{53}  &  \ding{53}  &  \ding{53} &  \small{Spitzer}  &  S12  	\\
M\,1$-$69  &  038.7$-$03.3  &  ORD  &  \ding{53}  &  \ding{53}  &  \ding{53}  &  \ding{53}  &  \checkmark &  \small{Spitzer}  &  S12  	\\
M\,1$-$71  &  055.5$-$00.5  &  CRD  &  \checkmark  &  \checkmark  &  \checkmark  &  \ding{53}  &  \ding{53} &  \small{Spitzer}  &  S12 	\\
M\,2$-$2  &  147.8$+$04.1  &  F  &  \ding{53}  &  \ding{53}  &  \ding{53}  &  \ding{53}  &  \ding{53} &  \small{Spitzer}  &  t.w. 	\\
M\,2$-$5  &  351.2$+$05.2  &  MD  &  \checkmark  &  \ding{53}  &  \ding{53}  &  \ding{53}  &  \checkmark &  \small{Spitzer}  &  G08 	\\
M\,2$-$10  &  358.3$+$01.2  &  MD  &  \checkmark  &  \ding{53}  &  \ding{53}  &  \ding{53}  &  \checkmark  &  \small{Spitzer}  &  S12	\\
M\,2$-$14  &  004.1$-$03.8  &  ORD  &  \ding{53}  &  \ding{53}  &  \ding{53}  &  \ding{53}  &  \checkmark &  \small{Spitzer}  &  S12  	\\
M\,2$-$20  &  000.4$-$01.9  & $\cdots$  &  $\cdots$  &  \ding{53}  &  \ding{53}  &  \ding{53}  &  \ding{53} &  \small{Spitzer}  &  poor quality 	\\
M\,2$-$23  &  002.2$-$02.7  &  ORD  &  \ding{53}  &  \ding{53}  &  \ding{53}  &  \checkmark  &  \ding{53} &  \small{Spitzer}  &  S12 	\\
M\,2$-$27  &  359.9$-$04.5  &  MD  &  \checkmark  &  \ding{53}  &  \ding{53}  &  \ding{53}  &  \checkmark &  \small{Spitzer}  &  PC09 	\\
M\,2$-$31  &  006.0$-$03.6  &  MD  &  \checkmark  &  \ding{53}  &  \ding{53}  &  \ding{53}  &  \checkmark &  \small{Spitzer}  &  S12 	\\
M\,2$-$39  &  008.1$-$04.7  &  ORD  &  \ding{53}  &  \ding{53}  &  \ding{53}  &  \checkmark  &  \checkmark &  \small{Spitzer}  &  S12 	\\
M\,2$-$42  &  008.2$-$04.8  &  F  &  \ding{53}  &  \ding{53}  &  \ding{53}  &  \ding{53}  &  \ding{53} &  \small{Spitzer}  &  S12 	\\
M\,2$-$43  &  027.6$+$04.2  & CRD? & $\cdots$ &  \checkmark? &    \ding{53} &  \ding{53}  & \ding{53}  &  \small{IRAS/Spitzer}  & Z90 	\\
M\,2$-$50  &  097.6$-$02.4  &  ORD?  &  \ding{53}  &  \ding{53}  &  \ding{53}  &  \checkmark?  &  \ding{53} &  \small{Spitzer}  &  S12 	\\
M\,2$-$51  &  103.2$+$00.6  & ORD?  &  \ding{53}?  &  \ding{53}  &  \ding{53}  &  \ding{53}  &  \checkmark &  \small{Spitzer}  &  poor quality 	\\
M\,3$-$8  &  051.0$+$02.8  &  ORD  &  \ding{53}  &  \ding{53}  &  \ding{53}  &  \checkmark  &  \checkmark &  \small{Spitzer}  &  S12 	\\
M\,3$-$10  &  358.2$+$03.6  &  ORD  &  \ding{53}  &  \ding{53}  &  \ding{53}  &  \checkmark  &  \checkmark &  \small{Spitzer}  &  S12 	\\
M\,3$-$13  &  005.2$+$04.2  &  MD  &  \checkmark  &  \ding{53}  &  \ding{53}  &  \checkmark?  &  \checkmark &  \small{Spitzer}  &  PC09 	\\
M\,3$-$15  &  006.8$+$04.1  &  MD  &  \checkmark  &  \ding{53}  &  \ding{53}  &  \ding{53}  &  \checkmark &  \small{Spitzer}  &  PC09 	\\
M\,3$-$23  &  052.9$-$02.7  &  F &  \ding{53}  &  \ding{53}  &  \ding{53}  &  \ding{53}  &  \ding{53}? &  \small{Spitzer}  &  S12	\\
M\,3$-$25  &  019.7$+$03.2  &  MD  &  \checkmark  &  \ding{53}  &  \ding{53}  &  \checkmark  &  \checkmark &  \small{Spitzer}  &  S12 	\\
M\,3$-$27  &  043.3$+$11.6  &  MD  &  \checkmark  &  \ding{53}  &  \ding{53}  &  \checkmark  &  \checkmark &  \small{Spitzer}  &  S12  	\\
M\,3$-$35  &  071.6$-$02.3  & $\cdots$  &  \ding{53}?  &  \ding{53}  &  \ding{53}  &  \checkmark  &  \ding{53} &  \small{IRAS/Spitzer}  & Z90 	\\
M\,3$-$38  &  356.9$+$04.4  &  MD  &  \checkmark  &  \ding{53}  &  \ding{53}  &  \checkmark  &  \checkmark &  \small{Spitzer}  &  PC09 	\\
M\,3$-$39  &  004.0$-$11.1  & $\cdots$  &  \checkmark  &  $\cdots$  &  $\cdots$  &  $\cdots$  &  $\cdots$ &  \small{Spitzer}  &  poor quality 	\\
M\,3$-$40  &  358.7$+$05.2  &  MD  &  \checkmark  &  \ding{53}  &  \ding{53}  &  \ding{53}  &  \checkmark &  \small{Spitzer}  &  S12 	\\
M\,3$-$44  &  359.3$-$01.8  &  MD  &  \checkmark  &  \ding{53}  &  \ding{53}  &  \ding{53}  &  \checkmark &  \small{Spitzer}  &  PC09 	\\
M\,3$-$47  &  000.3$-$02.8  &  F  &  \ding{53}  &  \ding{53}  &  \ding{53}  &  \ding{53}  &  \ding{53} &  \small{Spitzer}  &  S12 	\\
M\,3$-$50  &  357.1$-$06.1  &  F  &  \ding{53}  &  \ding{53}  &  \ding{53}  &  \ding{53}  &  \ding{53} &  \small{Spitzer}  &  S12 	\\
M\,3$-$54  &  018.6$-$02.2  &  F  &  \ding{53}  &  \ding{53}  &  \ding{53}  &  \ding{53}  &  \ding{53} &  \small{Spitzer}  &  S12 	\\
M\,4$-$5  &  000.7$+$03.2  &  ORD  &  \ding{53}  &  \ding{53}  &  \ding{53}  &  \ding{53}  &  \checkmark &  \small{Spitzer}  &  G08 	\\
M\,4$-$6  &  358.6$+$01.8  &  MD?  &  \checkmark?  &  \ding{53}  &  \ding{53}  &  \ding{53}  &  \checkmark &  \small{Spitzer}  &  S12 	\\
M\,4$-$10  &  019.2$-$02.2  &  ORD  &  \ding{53}  &  \ding{53}  &  \ding{53}  &  \checkmark  &  \checkmark &  \small{Spitzer}  &  S12 	\\
M\,4$-$18  &  013.4$-$03.9  &  CRD  &  \checkmark  &  \checkmark  &  \ding{53}  &  \ding{53}  &  \ding{53} &  \small{Spitzer}  &  S12 	\\
M\,57  &  063.1$+$13.9  & $\cdots$  &  $\cdots$  &  \ding{53}  &  $\cdots$  &  \ding{53}  &  $\cdots$ &  \small{ISO}  & P99 	\\
MA\,13  &  023.9$+$01.2  &  MD  &  \checkmark  &  \ding{53}  &  \ding{53}  &  \checkmark  &  \checkmark &  \small{Spitzer}  &  S12 	\\
MaC\,1$-$2  &  107.4$-$00.6  & $\cdots$  &  \checkmark  &  $\cdots$  &  $\cdots$  &  $\cdots$  &  $\cdots$ &  \small{Spitzer}  &  S12 	\\
MaC\,1$-$11  &  008.6$-$02.6  &  ORD?  &  \ding{53}?  &  \ding{53}  &  \ding{53}  &  \ding{53}  &  \checkmark? &  \small{Spitzer}  &  S12 	\\
MaC\,2$-$1  &  205.8$-$26.7  &  CRD  &  \ding{53}  &  \checkmark  &  \checkmark  &  \ding{53}  &  \ding{53} &  \small{Spitzer}  &  S12 	\\
MR\,22  &  284.1$-$00.7  & $\cdots$ &  $\cdots$  & $\cdots$ &  \ding{53}  & $\cdots$ &  \checkmark  &  \small{ISO}  & M02 	\\
Mz\,3  &  331.7$-$01.0  & $\cdots$  &  $\cdots$  &  \ding{53}  &  \ding{53}  &  \checkmark  &  \checkmark &  \small{ISO}  & BS05 	\\
NGC\,40  &  120.0$+$09.8  &  CRD  &  \checkmark  &  \checkmark  &  \checkmark  &  \ding{53}  &  \ding{53} &  \small{Spitzer/ISO}  &  H01, DI14	\\
NGC\,246  &  118.8$-$74.7  &  F  &  \ding{53}  &  \ding{53}  &  \ding{53}  &  \ding{53}  &  \ding{53} &  \small{Spitzer}  &  t.w. 	\\
NGC\,650  &  311.1$+$03.4  &  F?  &  \ding{53}?  &  \ding{53}  &  \ding{53}  &  \ding{53}  &  \ding{53} &  \small{Spitzer}  &  S12  	\\
NGC\,1501  &  144.1$+$06.1  & $\cdots$  &  $\cdots$  &  \ding{53}  &  \ding{53}  &  \ding{53}  &  \ding{53} &  \small{Spitzer}  &  t.w. 	\\
NGC\,1535  &  206.4$-$40.5  &  CRD?  &  \ding{53}  &  \ding{53}  &  \checkmark?  &  \ding{53}  &  \ding{53} &  \small{Spitzer}  &  S12	\\
NGC\,2346  &  215.6$+$03.6  & $\cdots$  &  \checkmark  &  $\cdots$  &  $\cdots$  &  $\cdots$  &  $\cdots$ &  \small{Spitzer}  &  t.w. 	\\
NGC\,2392  &  197.8$+$17.3  &  F  &  \ding{53}  &  \ding{53}  &  \ding{53}  &  \ding{53}  &  \ding{53} &  \small{Spitzer}  &  DI14 	\\
NGC\,2440  &  234.8$+$02.4  &  oPAH  &  \checkmark  &  \ding{53}  &  \ding{53}  &  \ding{53}  &  \ding{53} &  \small{ISO}  & BS02 	\\
NGC\,2792  &  265.7$+$04.1  &  F  &  \ding{53}  &  \ding{53}  &  \ding{53}  &  \ding{53}  &  \ding{53} &  \small{Spitzer}  &  t.w. 	\\
NGC\,2867  &  278.1$-$05.9  &  MD  &  \checkmark  &  \ding{53}  &  \ding{53}  &  \ding{53}  &  \checkmark &  \small{Spitzer}  &  PC09 	\\
NGC\,3132  &  272.1$+$12.3  &  ORD  &  \ding{53}  &  \ding{53}  &  \ding{53}  &  \ding{53}  &  \checkmark &  \small{Spitzer}  &  PC09 	\\
NGC\,3242  &  261.0$+$32.0  & CRD?  &  \ding{53}  &  \ding{53}  &  \checkmark?  &  \ding{53}  &  \ding{53} &  \small{Spitzer}  & DI14 	\\
NGC\,3918  &  294.6$+$04.7  & oPAH? &  \checkmark?  &  \ding{53}  & \ding{53} &  \ding{53}  & \ding{53} &  \small{Spitzer}  &  G08  	\\
NGC\,5315  &  309.1$-$04.3  & ORD? &  \ding{53}  &  \ding{53}? &  \ding{53}  &  \ding{53}  & \checkmark? &  \small{IRAS/Spizer}  & P86 	\\
NGC\,5882  &  327.8$+$10.0  & $\cdots$  &  \ding{53}  &  \ding{53}  &  $\cdots$  &  \ding{53}  &  $\cdots$  &  \small{IRAS}  & P86 	\\
NGC\,6072  &  343.4$+$11.9  &  F  &  \ding{53}  &  \ding{53}  &  \ding{53}  &  \ding{53}  &  \ding{53} &  \small{Spitzer}  &  S12  	\\
NGC\,6153  &  341.8$+$05.4  & $\cdots$  &  $\cdots$  &  \ding{53}  &  $\cdots$  &  \ding{53}  &  \checkmark &  \small{ISO/IRAS}  & P86, BS05 	\\
NGC\,6210  &  043.1$+$37.7  &  ORD  &  \ding{53}  &  \ding{53}  &  $\cdots$  &  \ding{53}  &  \checkmark &  \small{Spitzer}  &  DI14	\\
NGC\,6302  &  349.5$+$01.0  &  MD  &  \checkmark  &  \ding{53}  &  \ding{53}  &  \checkmark?  &  \checkmark &  \small{ISO}  &  M02, BS05 	\\
NGC\,6309  &  009.6$+$14.8  & $\cdots$  &  \ding{53}  &  \ding{53}  &  $\cdots$  &  \ding{53}  &  $\cdots$ &  \small{IRAS}  & R83 	\\
NGC\,6369  &  002.4$+$05.8  &  CRD  &  \ding{53}  &  \ding{53}  &  \checkmark  &  \ding{53}  &  \ding{53} &  \small{ISO}  & H01, H02b 	\\
NGC\,6439  &  011.0$+$05.8  &  ORD  &  \ding{53}  &  \ding{53}  &  \ding{53}  &  \ding{53}  &  \checkmark &  \small{Spitzer}  &  DI14 	\\
NGC\,6537  &  010.0$+$00.7  & $\cdots$  &  $\cdots$  &  \ding{53}  &  \ding{53}  &  \ding{53}  &  \checkmark &  \small{ISO/IRAS}  & P86, M02 	\\
NGC\,6543  &  096.4$+$29.9  & $\cdots$  &  $\cdots$  &  \ding{53}  &  \ding{53}  &  \ding{53}  &  \ding{53} &  \small{ISO/IRAS}  &  P99, BS05 	\\
NGC\,6567  &  011.7$-$06.6  &  CRD? &  \checkmark?  &  \checkmark? &  \checkmark?  &  \ding{53}  &  \ding{53} &  \small{Spitzer}  &  S12 	\\
NGC\,6572  &  034.6$+$11.8  & $\cdots$  &  \ding{53}?  &  \ding{53}  &  $\cdots$  &  \checkmark  &  $\cdots$ &  \small{IRAS}  & P86 	\\
NGC\,6644  &  008.3$-$07.3  &  CRD  &  \checkmark  &  \checkmark  &  \checkmark  &  \ding{53}  &  \ding{53} &  \small{Spitzer}  &  PC09 	\\
NGC\,6741  &  033.8$-$02.6  & $\cdots$ & $\cdots$ &  \ding{53}  &  $\cdots$  &  \ding{53}  &  $\cdots$  &  \small{Spitzer}  & DI14 	\\
NGC\,6790  &  037.8$-$06.3  & $\cdots$ &  \ding{53}?  &  \ding{53}? &  $\cdots$  &  \ding{53}  &  $\cdots$ &  \small{IRAS}  & P86 	\\
NGC\,6807  &  042.9$-$06.9  &  ORD  &  \ding{53}  &  \ding{53}  &  \ding{53}  &  \checkmark  &  \checkmark &  \small{Spitzer}  &  S12 	\\
NGC\,6826  &  083.5$-$12.7  &  CRD  &  $\cdots$  &  \ding{53}  &  \checkmark  &  \ding{53}  &  \ding{53} &  \small{Spitzer}  & DI14 	\\
NGC\,6852  &  042.5$-$14.5  &  F  &  \ding{53}  &  \ding{53}  &  \ding{53}  &  \ding{53}  &  \ding{53} &  \small{Spitzer}  &  t.w. 	\\
NGC\,6881  &  074.5$+$02.1  &  oPAH  &  \checkmark  &  \ding{53}  &  \ding{53}  &  \ding{53}  &  \ding{53} &  \small{Spitzer}  &  PC09 	\\
NGC\,6884  &  082.1$+$07.0  & $\cdots$  &  \ding{53}  &  \ding{53}?  &  $\cdots$  &  \ding{53}?  &  $\cdots$ &  \small{IRAS}  & R86 	\\
NGC\,6886  &  060.1$-$07.7  & $\cdots$  &  \ding{53}  &  $\cdots$  &  $\cdots$  &  $\cdots$  &  $\cdots$ &  \small{Spitzer}  &  t.w. 	\\
NGC\,7008  &  093.4$+$05.4  &  F  &  \ding{53}  &  \ding{53}  &  \ding{53}  &  \ding{53}  &  \ding{53} &  \small{Spitzer}  &  t.w. 	\\
NGC\,7009  &  037.7$-$34.5  & $\cdots$  &  \ding{53}?  &  \ding{53}  &  $\cdots$  &  \ding{53}  &  $\cdots$ &  \small{IRAS/Spizer}  & P86 	\\
NGC\,7026  &  089.0$+$00.3  & MD  &  \checkmark  &  \ding{53}  & \ding{53}  &  \ding{53}  &  \checkmark  &  \small{Spitzer}  & DI14 	\\
NGC\,7027  &  084.9$-$03.4  &  CRD  &  \checkmark  &  \checkmark  &  \checkmark  &  \ding{53}  &  \ding{53} &  \small{ISO}  &  H01, BS05  	\\
NGC\,7094  &  066.7$-$28.2  &  F  &  \ding{53}  &  \ding{53}  &  \ding{53}  &  \ding{53}  &  \ding{53} &  \small{Spitzer}  &  t.w. 	\\
NGC\,7354  &  107.8$+$02.3  & $\cdots$  &  \ding{53}  &  \ding{53}  &  $\cdots$  &  \ding{53}  &  $\cdots$ &  \small{IRAS}  & P86 	\\
NGC\,7662  &  106.5$-$17.6  & $\cdots$  &  \ding{53}  &  \ding{53}  &  $\cdots$  &  \ding{53}  &  $\cdots$ &  \small{Spitzer}  & DI14 	\\
PB\,2  &  263.0$-$05.5  &  CRD  &  \ding{53}  &  \checkmark  &  \checkmark  &  \ding{53}  &  \ding{53} &  \small{Spitzer}  &  S12 	\\
PBOZ\,24  &  002.1$+$03.3  &  ORD  &  \ding{53}  &  \ding{53}  &  \ding{53}  &  \ding{53}  &  \checkmark &  \small{Spitzer}  &  G08 	\\
PBOZ\,26  &  285.4$+$02.2  &  F  &  \ding{53}  &  \ding{53}  &  \ding{53}  &  \ding{53}  &  \ding{53} &  \small{Spitzer}  &  S12  	\\
Pe\,1$-$1 &  285.4$+$01.5  &  CRD  &  \checkmark  &  \ding{53}  &  \checkmark  &  \ding{53}  &  \ding{53} &  \small{Spitzer}  &  S12  	\\
Pe\,2$-$8  &  322.4$-$00.1  & ORD?  &  \ding{53}  &  \ding{53}  &  $\cdots$  &  \checkmark  &  $\cdots$ &  \small{IRAS}  & P86 	\\
Pe\,2$-$10  &  006.8$+$02.0  &  ORD  &  \ding{53}  &  \ding{53}  &  \ding{53}  &  \ding{53}  &  \checkmark &  \small{Spitzer}  &  S12 	\\
Pe\,2$-$15  &  026.0$-$01.8  &  F  &  \ding{53}  &  \ding{53}  &  \ding{53}  &  \ding{53}  &  \ding{53} &  \small{Spitzer}  &  S12  	\\
PMR\,1  &  272.8$+$01.0  & oPAH?  &  \ding{53}?  &  \ding{53}  &  \ding{53}  &  \ding{53}  &  \ding{53} &  \small{Spitzer}  &  PC09 	\\
PMR\,2  &  291.3$+$08.4  &  F  &  \ding{53}  &  \ding{53}  &  \ding{53}  &  \ding{53}  &  \ding{53} &  \small{Spitzer}  &  t.w. 	\\
PMR\,4  &  014.2$+$03.8  &  F  &  \ding{53}  &  \ding{53}  &  \ding{53}  &  \ding{53}  &  \ding{53} &  \small{Spitzer}  &  t.w. 	\\
Sa\,1$-$5  &  340.9$-$04.6  &  F  &  \ding{53}  &  \ding{53}  &  \ding{53}  &  \ding{53}  &  \ding{53} &  \small{Spitzer}  &  S12 	\\
Sa\,2$-$6  &  258.0$-$15.7  & ORD? &  \ding{53}? &  \ding{53}  &  \ding{53}  &  \ding{53}  &  \checkmark &  \small{Spitzer}  &  S12  	\\
Sa\,2$-$157  &  325.8$-$12.8  &  ORD  &  \ding{53}  &  \ding{53}  &  \ding{53}  &  \checkmark  &  \ding{53} &  \small{Spitzer}  &  S12 	\\
Sa\,2$-$208  &  344.4$-$06.1  &  F  &  \ding{53}  &  \ding{53}  &  \ding{53}  &  \ding{53}  &  \ding{53} &  \small{Spitzer}  &  S12 	\\
Sa\,3$-$90  &  000.8$-$01.5  &  MD  &  \checkmark  &  \ding{53}  &  \ding{53}  &  \ding{53}  &  \checkmark &  \small{Spitzer}  &  S12 	\\
Sa\,3$-$134  &  016.9$-$02.0  &  MD  &  \checkmark  &  \ding{53}  &  \ding{53}  &  \ding{53}  &  \checkmark &  \small{Spitzer}  &  S12 	\\
Sp\,4$-$1  &  068.7$$+$$14.8  &  oPAH  &  \checkmark  &  \ding{53}  &  \ding{53}  &  \ding{53}  &  \ding{53} &  \small{Spitzer}  &  S12 	\\
SwSt\,1  &  001.5$-$06.7  & $\cdots$ &  \ding{53}  &  \ding{53}  &  $\cdots$  &  \checkmark  &  $\cdots$  &  \small{IRAS/Spitzer}  & Z90 	\\
Th\,3$-$6  &  354.9$$+$$03.5  &  MD  &  \checkmark  &  \ding{53}  &  \ding{53}  &  \ding{53}  &  \checkmark &  \small{Spitzer}  &  S12 	\\
Th\,3$-$10  &  355.9$$+$$02.7  &  MD  &  \checkmark  &  \ding{53}  &  \ding{53}  &  \ding{53}  &  \checkmark &  \small{Spitzer}  &  S12 	\\
Th\,3$-$12  &  356.8$$+$$03.3  &  MD  &  \checkmark  &  \ding{53}  &  \ding{53}  &  \ding{53}  &  \checkmark &  \small{Spitzer}  &  S12 	\\
Th\,3$-$24  &  357.1$$+$$01.9  &  F  &  \ding{53}  &  \ding{53}  &  \ding{53}  &  \ding{53}  &  \ding{53} &  \small{Spitzer}  &  S12 	\\
Th\,3$-$55  &  356.5$$+$$01.5  &  ORD  &  \ding{53}  &  \ding{53}  &  \ding{53}  &  \ding{53}  &  \checkmark &  \small{Spitzer}  &  S12 	\\
Th\,4$-$3  &  006.0$$+$$02.8  &  MD  &  \checkmark  &  \ding{53}  &  \ding{53}  &  \checkmark  &  \checkmark &  \small{Spitzer}  &  S12 	\\
Th\,4$-$6  &  009.3+04.1 &  ORD & \ding{53}  &  \ding{53}  &  \ding{53}  & \checkmark  &\checkmark  & \small{Spitzer}  &  S12  	\\
Vd\,1$-$2  &  345.0$$+$$04.3  &  ORD  &  \ding{53}  &  \ding{53}  &  \ding{53}  &  \checkmark  &  \ding{53} &  \small{Spitzer}  &  S12 	\\
Vd\,1$-$3  &  344.8$$+$$03.4  &  oPAH  &  \checkmark  &  \ding{53}  &  \ding{53}  &  \ding{53}  &  \ding{53} &  \small{Spitzer}  &  S12 	\\
Vd\,1$-$5  &  344.4$$+$$02.8  &  oPAH?  &  \checkmark?  &  \ding{53}  &  \ding{53}  &  \ding{53}  &  \ding{53} &  \small{Spitzer}  &  S12 	\\
VV\,3  &  014.3$-$05.5  &  CRD  &  \checkmark  &  \checkmark  &  \checkmark  &  \ding{53}  &  \ding{53} &  \small{Spitzer}  &  S12  	\\
Vy\,2$-$2  &  045.4$-$02.7  & $\cdots$  &  $\cdots$  &  \ding{53}  &  $\cdots$  &  \checkmark  &  $\cdots$ &  \small{ISO/Spitzer}  &  P99, H02a	\\
\hline
\\
\end{longtable}
\vspace*{-0.8cm}
\begin{center}
	\begin{minipage}{160mm}
	References for the published spectra: B83: \citet{1983IAUS..103..105B},  BS02: \citet{2002A&A...387..301B}, BS05: \citet{2005A&A...431..523B}, DI14: \citet{Delgado2014}, G08: \citet{Gutenskunst2008}, H01: \citet{2001A&A...378L..41H}, H02a:  \citet{2002A&A...390..533H}, H02b:  \citet{2002A&A...393L.103H}, M02: \citet{2002A&A...382..184M}, O14: \citet{Otsuka2014}, P84 \citet{1984ApJ...278L..33P}, P86: \citet{1986A&A...161..363P}, P99: \citet{1999A&A...351..201P}, PC09: \cite{PereaCalderon2009}, R83: \citet{1983MNRAS.204.1017R}, R86: \citet{1986MNRAS.221...63R}, S12: \citet{Stanghellini2012}, t.w.: this work, W98: \citet{Waters1998}, Z90: \citet{1990A&A...237..479Z}.
	\end{minipage}
\end{center}
\clearpage
\section{Objects with different assigned type}\label{app:differences}

We present in Table~\ref{tab:different_clasification_S12_GH14} the list of objects in which our classification differs from those of \cite{Stanghellini2012} and \cite{GarciaHernandez2014}. We have used the features that these authors identify to translate their classification into our classification scheme. The differences are related to the presence or absence of the dust features listed in the last column of 
Table~\ref{tab:different_clasification_S12_GH14}.

	\begin{table}
    	\centering
    	\caption{Objects where our classification differs from those of \citet[][S12]{Stanghellini2012} or \citet[][GH14]{GarciaHernandez2014}.}
    	\begin{tabular}{llllll}
      		\hline
       		Name & PN\,G & S12 & GH14 & This work & Feature \\
        	\hline
        	Al\,2-E & 359.3+03.6 & F & F & ORD & Crystalline silicates \\
        	H\,1-33 & 355.7$-$03.0  & oPAH & oPAH & MD & Crystalline silicates \\
        	H\,2-27 & 356.5$-$03.6 & F & $\cdots$ & ORD & Crystalline silicates \\
        	H\,2-39 & 002.9$-$03.9  & ORD & ORD & F & Crystalline silicates \\     	
        	Hen\,2-63 & 289.8+07.7 & F & F & oPAH & PAHs \\
        	Hen\,2-186 & 336.3$-$05.6  & oPAH & oPAH & MD & Crystalline silicates \\
        	Hen\,2-440 & 060.5+01.8  & ORD & ORD & MD & PAHs \\
        	IC\,4776 & 002.0$-$13.4  & MD & MD & ORD & PAHs \\
        	K\,3-15 & 041.8+04.4 & CRD & oPAH & CRD & SiC \\
        	K\,3-29 & 048.1+01.1 & oPAH & $\cdots$ & CRD & 30 $\mu$m \\
        	K\,4-8 & 025.3$-$04.6 & ORD & ORD & F & Crystalline silicates \\ 
        	M\,1-69 & 038.7$-$03.3 & F & F & ORD & Crystalline silicates \\       	
        	M\,2-10 & 354.2+04.3  & oPAH & MD & MD & Crystalline silicates \\
        	M\,2-14 & 003.6+03.1  & $\cdots$ & MD & ORD & PAHs \\
        	M\,3-8 & 358.2+04.2  & $\cdots$ & MD & ORD & PAHs \\
        	NGC\,6072 & 343.4+11.9 & $\cdots$ & F & ORD & Crystalline silicates \\        	
        	Pe\,1-1 & 285.4+01.5 & oPAH & oPAH & CRD & 30 $\mu$m \\
		Th\,4$-$6  &  009.3+04.1 &  ORD & MD & ORD & PAHs \\
        	VV\,3 & 014.3$-$05.5 & oPAH & oPAH & CRD & SiC \\
       		\hline
    	\end{tabular}  
        \label{tab:different_clasification_S12_GH14}
	\end{table}
	

\bsp	
\label{lastpage}
\end{document}